\documentclass[aps,rmp]{revtex4}
\usepackage[a4paper,top=2.00cm,bottom=2.00cm,left=2.00cm,right=2.00cm]{geometry}
\usepackage{amssymb}
\usepackage{graphicx}
\newcommand{\ie}{{\it i.e.\ }}
\newcommand{\eg}{{\it e.g.\ }}
\newcommand{\et}{{\it et al.\ }}
\newcommand{\be}{\begin{equation}}
\newcommand{\ee}{\end{equation}}
\newcommand{\ber}{\begin{eqnarray}}
\newcommand{\eer}{\end{eqnarray}}
\begin{document}
%~~~~~~~~~~~~~~~~~~~~~~~~~~~~~~~~~~~~~~~~~~~~~~~~~~~~~~~~~~~~~~~~~~~~~~~~~~~~~~~~~
\pagenumbering{arabic}
\setcounter{page}{106}
\title{Fermi-Liquid Theory for Unconventional Superconductors
\\
\vspace{-55mm}{\footnotesize
Published in ``Strongly Correlated Electronic Materials - The Los Alamos Symposium 1993'',
K. Bedell, Z. Wang, D. Meltzer, A. Balatsky and E. Abrahams, eds., 
Addison-Wesley Publishing Co., New York (1994). ISBN 0-2-1-40930-5.}
\vspace*{50 mm}
}
\author{J.A. Sauls}
\affiliation{Department of Physics \& Astronomy, Northwestern University \\ 
             Evanston, IL 60208, USA}

\begin{abstract}
\vspace*{2.5cm}
\begin{minipage}{0.8\textwidth}
Fermi liquid theory is used to generate the Ginzburg-Landau free energy functionals for unconventional superconductors belonging to various representations. The parameters defining the GL functional depend on Fermi surface anisotropy, impurity scattering and the symmetry class of the pairing interaction. As applications I consider the basic models for the superconducting phases of UPt$_3$. Two predictions of Fermi liquid theory for the two-dimensional representations of the hexagonal symmetry group are (i) the zero-field equilibrium state exhibits spontaneously broken time-reversal symmetry, and (ii) the gradient energies for the different 2D representations, although described by a similar GL functionals, are particularly sensitive to the orbital symmetry of the pairing state and Fermi surface anisotropy.
\end{minipage}
\end{abstract}
\maketitle
%~~~~~~~~~~~~~~~~~~~~~~~~~~~~~~~~~~~~~~~~~~~~~~~~~~~~~~~~~~~~~~~~~~~~~~~~~~~~~~~~~

\section{Introduction}

The heavy fermions and cuprates represent classes of strongly
correlated metals in which the mechanism responsible for
superconductivity is probably electronic in origin. In the heavy
fermions the metallic state develops below a coherence temperature
$T^{*}\sim 10\,K$, well below the Debye temperature. Heavy electronic
quasiparticles have Fermi velocities comparable to the sound velocity.
Thus, Migdal's expansion in $c_s/v_f$, which is the basis for the
theory of conventional strong-coupling superconductors, breaks down.
Overscreening of the Coulomb interaction by ions is suppressed, leaving
strong, short-range Coulomb interactions to dominate the effective
interaction between heavy electron quasiparticles.  Phonon-mediated
superconductivity is possible, but the breakdown of the Migdal
expansion favors electronically driven superconductivity, or requires
an electron-phonon coupling that is strongly enhanced by electronic
correlations.$^{1}$ Similarly, for high $T_c$ materials there is no
`theorem' precluding superconductivity at $T_c\sim\,100\,K$ from the
electron-phonon interaction, but the existence of high transition
temperatures has often been given as reason enough for abandoning
phonons as the mechanism for superconductivity in the cuprates.$^{2}$
Indeed much of the theoretical effort in trying to understand the
origin of superconductivity in the cuprates is devoted to models based
on electronic mechanisms of superconductivity ({\it c.f.} these
proceedings).

An important consequence of many electronic models for
superconductivity is that they often favor a superconducting state with
an {\it unconventional} order parameter, {\ie} a pair amplitude that
spontaneously breaks one or more symmetries of the metallic phase in
addition to gauge symmetry. An order parameter with lower symmetry
than the normal metallic phase can lead to dramatic effects on the
superconductivity and low-lying excitation spectrum. The gap in the
quasiparticle excitation spectrum may vanish at points or lines on the
Fermi surface, reflecting a particular broken symmetry. Such nodes are
robust features of the excitation spectrum that lead to power law
behavior as $T\rightarrow 0$ for Fermi-surface averaged thermodynamic
and transport properties. These average properties have been the focus
of a considerable experimental and theoretical effort in order to
identify the nature of the order parameter, particularly in the heavy
fermion superconductors.$^{3}$

However, the most striking differences between conventional and
unconventional superconductors are those properties that exhibit a
broken symmetry, or reveal the residual symmetry, of the order
parameter. Examples of properties are (i) spontaneously broken
rotational symmetries exhibited by the anisotropy of the penetration
depth tensor in tetragonal, hexagonal and cubic lattice structures,
(ii) new types of vortices also reflecting broken symmetries of the
ordered phases, (iii) new collective modes of the order parameter that
are distinct from the phase and amplitude mode in conventional
superconductors, (iv) sensitivity of the superconductivity to
non-magnetic impurity and surface scattering, (v) anomalous Josephson
effects associated of combined gauge-rotation symmetries of the order
parameter, and (vi) complex phase diagrams associated with
superconducting phases with different residual symmetries. Indeed, the
strongest evidence for unconventional superconductivity in metals is
the multiple superconducting phases of $(U,Th)Be_{13}$ and
$UPt_3$.$^{4,5}$ For further discussion of these issues see the
reviews by Gorkov,$^{6}$ Sigrist and Ueda,$^{7}$ and Muzikar,
{\et}$^{8}$

At this workshop I summarized experimental evidence for, and
theoreticial interpretations of, unconventional superconductivity in
the heavy fermion superconductor. I focussed on UPt$_3$ because its
complex phase diagram leads to strong constraints on the symmetry of
superconducting order parameter. Another reason is that
superconductivity onsets at $T_c\sim 0.5\,K\ll T^{*}\sim 10\,K$, in the
well-developed Fermi-liquid. This separation in energy scales is
important for formulating a theory of superconductivity in strongly
correlated Fermion systems with predictive power. In this article I
discuss the theoretical framework that supports the phenomenology on
unconventional superconductivity in UPt$_3$ that I presented at the
workshop.

The heavy fermion metals, high-$T_c$ cuprates and liquid $^3$He are
strongly correlated fermion systems for which we do not have a
practical first-principals theory of superconductivity and
superfluidity. However, we have powerful and successful
phenomenological theories for superconductivity in such systems: the
Ginzburg-Landau theory and the Fermi-liquid theory of
superconductivity. The full power of these phenomenological theories
is realized in there connection to one another and to the experimental
data on low energy phenomena. The Fermi-liquid theory of
superconductivity reduces to GL theory in the limit $T\rightarrow T_c$,
with specific predictions for the material parameters in terms of those
of Fermi-liquid theory, {\eg} the Fermi surface, Fermi velocity,
electronic density of states, electronic interactions, mean-free path,
etc. Fermi-liquid theory then extends beyond the range of the GL
theory, to low-temperatures, higher fields, and shorter wavelengths. In
what follows I develop the Fermi-liquid theory of strongly correlated
metals and derive the GL functionals for superconductors with an
unconventional order parameter.  As applications I examine the basic
models for the superconducting phases of UPt$_3$.$^{9}$

In Section II I summarize the GL theory for superconductors with an
unconventional order parameter, and construct free energy functionals
for several models of superconductivity in the heavy fermion
superconductors. Section III deals with the formulation of Fermi-liquid theory for
strongly correlated metals from microscopic theory. I start with the
stationary free-energy functional of Luttinger and Ward,$^{10}$ then
formulate the Fermi-liquid theory in terms of an expansion in the
low-energy, long-wavelength parameters represented by $\mbox{\sl small} \sim
T/T^*$, $\Delta/T^*$, $1/k_f\xi_0$, etc. For inhomogeneous
superconducting states the central equation of the Fermi-liquid theory
is the quasiclassical transport equation, which generalizes the
Boltzmann-Landau transport equation to superconducting Fermi liquids.
External fields couple to the quasiparticle excitations and enter the
transport equation through the self energy. The leading order
self-energies describe impurity scattering, electron-electron
interactions and the coupling to external magnetic fields. The
resulting stationary free energy functional, due to Rainer and
Serene,$^{11}$ combined with the leading order self energies, is the
basis for derivations the GL theory from Fermi-liquid theory.

Material parameters are calculated for the 2D representations that have
been discussed for UPt$_3$. An important prediction of the leading
order Fermi-liquid theory, for any of the 2D representations in
uniaxial superconductors, is that the equilibrium state in zero field
spontaneously breaks time-reversal symmetry. The result is robust; it
is independent of the details of the Fermi surface anisotropy, the
basis functions determining the anisotropic gap function, as well as
s-wave impurity scattering.

Extensions of the theory to low-temperature properties are also
discussed.  The linearized gap equation, with both diamagnetic and
paramagnetic contributions, for odd-parity superconductors with strong
spin-orbit coupling is examined. The gradient terms of the GL
free-energy functional are related to Fermi-surface averages of
products of the Fermi velocity and the pairing basis functions. These
coefficients differ substantially for the various 2D representations,
even though the phenomenological GL functionals of all the 2D
representations are formally the same. The effects of Fermi surface
anisotropy on the gradient coefficients are calculated for order
parameters belonging to the E$_{1u}$ and E$_{2u}$ representations of
$D_{6h}$ (appropriate for UPt$_3$). All of these results have
implications for models of the superconducting phases of UPt$_3$, and
perhaps other heavy Fermion superconductors.$^{9}$

\section{Ginzburg-Landau Theory}

Landau's theory of second-order phase transitions is very general; the
structure of the theory is determined principally by symmetry and a few
constraints on the parameters defining the theory. Its quantitative
accuracy relies on the size of the pair correlation length being large
compared to the atomic scale $k_f^{-1}$ {\ie} $1/k_f\xi_0\sim
T_c/T^{*}<<1$, which are well satisfied in most superconductors,
including many of the heavy fermions. The generality of Ginzburg-Landau
(GL) theory comes at the price of restricted predictive power; the GL
theory depends on phenomenological material parameters that must be
determined by comparison with experiment, or from a more fundamental
theory. Futhermore, different GL theories (based on order parameters
with different symmetry properties) can lead to similar phase diagrams
and thermodynamic properties. Nevertheless, GL theory is an essential
tool for interpreting the magnetic and thermodynamic properties of
superconductors, and it has been used extensively to examine the
possible phases for superconductors with an unconventional order
parameter.$^{12-19}$

I assume that heavy fermion superconductors, and possibly the cuprates,
are all described by an equal-time pairing amplitude of the form,
\be
f_{\alpha\beta}(\vec{k}_f)\sim
\left<a_{\vec{k}_f\alpha}a_{-\vec{k}_f\beta}\right>
\,,
\ee
where $\alpha,\beta$ are spin labels of the quasiparticles.
In conventional superconductors the pair amplitude has the form,
$f_{\alpha\beta}(\vec{k}_f)=f_0(\vec{k}_f)\,(i\sigma_y)_{\alpha\beta}\,$, 
describing pairs with total spin zero and a (complex) amplitude
$f_0(\vec{k}_f)$ that breaks gauge symmetry, but otherwise retains the
full symmetry of the normal metal. {\it Unconventional}
superconductivity occurs when the pair amplitude spontaneously breaks
additional symmetries of the normal metallic state, {\ie} if there
exists an operation $R\,\varepsilon\,{\bf G}$, other than a gauge
transformation, for which $R*f(R^{-1}*\vec{k}_f)\ne
f(\vec{k}_f)$; ${\bf G}$ is the space group of the normal
state, combined with time-reversal and the gauge group, ${\bf G} =
G_s\times {\cal T}\times U(1)$. I assume that translation symmetry of
the normal metal is unbroken at the superconducting transition, in
which case $G_s$ may be indentified with the symmetry group of
rotations and inversions. In the heavy fermion superconductors it is
generally assumed that the spin-orbit interaction is strong (on the
scale of $T_c$) so that only joint rotations of the orbital and spin
coordinates are symmetry operations.$^{20,12}$ In this
case, $G_s$ is identified with the point group, $G_p$. The labels for
the quasiparticle states near the Fermi level are not eigenvalues of
the spin operator for electrons. Nevertheless, in zero-field the
Kramers degeneracy guarantees that each $\vec{k}$ state is two-fold
degenerate, and thus, may be labeled by a pseudo-spin quantum number
$\alpha$, which can take on two possible values. Furthermore, the
degeneracy of each $\vec{k}$-state is lifted by a magnetic field, which
is described by a Zeeman energy that couples the magnetic field to the
pseudo-spin with an effective moment that in general depends on the
orientation of the magnetic field relative to crystal coordinates, and
possibly the wavevector $\vec{k}$. In the opposite limit of negligible
spin-orbit coupling the normal
state is separately invariant under rotations in spin space, so
$G_s=SO(3)_{spin}\times G_p$.

Fermion statistics of the quasiparticles requires the pair amplitude to
obey the anti-symmetry condition,
\be
f_{\alpha\beta}(\vec{k}_f)=-f_{\beta\alpha}(-\vec{k}_f)
\,.
\ee
If the normal metal has inversion symmetry, then ${\bf G}$ contains the
two-element subgroup $(1,{\cal C}_i)$. This is the case for nearly all
systems of interest. For example, UPt$_3$, which is hexagonal with an
inversion center, is described by the group ${\bf G} = D_{6h}\times
{\cal T}\times U(1)$ for strong spin-orbit coupling, while the layered
$CuO$ superconductors have tetragonal symmetry with a point group
$D_{4h}$, or are weakly orthorhombic. The pairing interaction which
drives the superconducting instability depends on the momenta and spins
of quasiparticle pairs $(\vec{k}_f\alpha,-\vec{k}_f\beta)$ and
$(\vec{k}_f'\alpha,-\vec{k}_f'\beta)$ and, and necessarily decomposes
into even- and odd-parity sectors,
\ber
V_{\alpha\beta;\gamma\rho}(\vec{k}_f,\vec{k}_f')
&=&
\sum_{\Gamma}^{even}\,V_{\Gamma}\,\sum_{i=1}^{d_{\Gamma}}\,
{\cal Y}_i^{(\Gamma)}(\vec{k}_f)
(i\sigma_y)_{\alpha\beta}\,
{\cal Y}_i^{(\Gamma)}(\vec{k}_f')^{*}
(i\sigma_y)_{\gamma\rho}
\nonumber\\
&+&
\sum_{\Gamma}^{odd}\,V_{\Gamma}\,\sum_{i=1}^{d_{\Gamma}}\,
\vec{{\cal Y}}_i^{(\Gamma)}(\vec{k}_f)
 \cdot(i\sigma_y\vec{\sigma})_{\alpha\beta}\,
\vec{{\cal Y}}_i^{(\Gamma)}(\vec{k}_f')^{*}
 \cdot(i\vec{\sigma}\sigma_y)_{\gamma\rho}
\,.
\eer

The interaction is invariant under the operations of the group ${\bf G}$ and decomposes into a sum
over invariant bilinear products of basis functions for each irreducible representation $\Gamma$ of
the point group, with both even- and odd-parity. The basis functions, $\{{\cal
Y}^{(\Gamma)}_{i}(\vec{k}_f)\}$, for the symmetry groups of the heavy fermion superconductors are
tabulated in Ref. 21. Representative basis functions for the group $D_{6h}$, appropriate for UPt$_3$
with strong spin-orbit coupling, are given in Table I.

%------------------- Table of Basis Funcions -----------------------
\bigskip
\centerline{Table I: Basis functions for $D_{6h}$}
\medskip 
\centerline{\vbox{\offinterlineskip\hrule
\halign{\vrule #\hfil& \vrule \quad #\hfil& \vrule\ \vrule\quad
#\hfil&\vrule \quad #\hfil\vrule\cr
$\Gamma$& Even parity &$\Gamma$ & Odd parity ($\vec{d}||\hat{z}$)\cr
\noalign{\hrule}
A$_{1g}$ & 1 & A$_{1u}$ & $k_z$  \cr
A$_{2g}$ & $Im(k_x+ik_y)^6$ & A$_{2u}$ & $k_z Im(k_x+ik_y)^6$  \cr
B$_{1g}$ & $k_z\,Im(k_x+ik_y)^3$ & B$_{1u}$ & $Im(k_x+ik_y)^3$ \cr
B$_{2g}$ & $k_z\,Re(k_x+ik_y)^3$ & B$_{2u}$ & $Re(k_x+ik_y)^3$ \cr
 $E_{1g}$ & $k_z{\matrix{k_x \cr k_y \cr}}$ &
E$_{1u}$ & $\left(\matrix{k_x \cr k_y \cr}\right)$ \cr
 E$_{2g}$ & $\left(\matrix{k_x^2-k_y^2 \cr 2k_xk_y \cr}\right)$ &
E$_{2u}$ & $k_z\left(\matrix{k_x^2-k_y^2 \cr 2k_xk_y \cr}\right)$ \cr}
\hrule}
}
%-------------------------------------------------------------------

The order parameter separates into even- and odd-parity sectors:
\be
f_{\alpha\beta}(\vec{k}_f)=
f(\vec{k}_f)\,(i\sigma_y)_{\alpha\beta}
+
\vec{f}(\vec{k}_f)\cdot(i\vec{\sigma}\sigma_y)_{\alpha\beta}
\,,
\ee
where the even-parity (singlet) and odd-parity (triplet) order parameters
have the general form,
\be
f(\vec{k}_f)=\sum_{\Gamma}^{even}\sum_i^{d_{\Gamma}}
\,\eta_{i}^{(\Gamma)}\,{\cal Y}^{(\Gamma)}_{i}(\vec{k}_f)
\quad , \quad
\vec{f}(\vec{k}_f)=\sum_{\Gamma}^{odd}\sum_i^{d_{\Gamma}}
\,\eta_{i}^{(\Gamma)}\,\vec{{\cal
Y}}^{(\Gamma)}_{i}(\vec{k}_f)
\,.
\ee
There is an interaction parameter $V_{\Gamma}$ for each irreducible
representation $\Gamma$. The superconducting instability is determined
by the irreducible representation $\Gamma^{*}$ with the highest
transition temperature, and barring accidental degeneracies, the order
parameter, at least near $T_c$, will belong to the representation
$\Gamma^{*}$.

\subsection{Free Energy Functionals}

In order to analyze the stability of the possible superconducting
states a free-energy functional of the order parameter is needed; for
temperatures close to the transition temperature this is the
Ginzburg-Landau functional. The GL functional is invariant under the
symmetry operations of the group ${\bf G}$ of the normal state, and is
stationary at the equilibrium values of $\{\eta_i^{(\Gamma)}\}$ and
equal to the thermodynamic potential. The general form of the GL
functional is constructed from a symmetry analysis of the
transformation properties of products of the order parameter and
gradients of the order parameter. The procedure is well known and has
been carried out for many of the possible realizations of
unconventional superconductivity.$^{7}$

The general form of the GL functional includes one quadratic invariant
for each irreducible representation,
\be
\Omega_{GL}=\sum_{\Gamma}^{irrep}\,\alpha_{\Gamma}(T)\,
\sum_i^{d_{\Gamma}}(\eta_i^{(\Gamma)}\eta_i^{(\Gamma)*}) + ...
\,.
\ee

The coefficients $\alpha_{\Gamma}(T)$ are material parameters that
depend on temperature and pressure. Above $T_c$ all the coefficients
$\alpha_{\Gamma}(T)>0$. The instability to the superconducting state is
the point at which one of the coefficients vanishes, {\eg}
$\alpha_{\Gamma^*}(T_c)=0$. Thus, near $T_c$
$\alpha_{\Gamma^*}(T)\simeq\alpha'(T-T_c)$ and $\alpha_{\Gamma}>0$ for
$\Gamma\ne\Gamma^*$. At $T_c$ the system is unstable to the development
of all the amplitudes $\{\eta_i^{(\Gamma^*)}\}$, however, the higher
order terms in the GL functional which stabilize the system, also
select the ground state order parameter from the manifold of degenerate
states at $T_c$. In most superconductors the instability is in the
even-parity, $A_{1g}$ channel. This is conventional superconductivity
in which only gauge symmetry is spontaneously broken. An instability in
any other channel is a particular realization of unconventional
superconductivity.

There are two representative classes of GL theories; (i) those based on
a {\it single} primary order parameter belonging to a higher
dimensional representation of ${\bf G}$, and (ii) models based on two
primary order parameters belonging to different irreducible
representations which are nearly degenerate. Both types of model have
been examined as models for the the multiple superconducting phases of
UPt$_3$ and $(U,Th)Be_{13}$. The specific applications of these GL
theories to the phase diagrams of UPt$_3$ and $UBe_{13}$ are discussed
in detail elsewhere.$^{7,9,16,18,19}$

\subsection{2D Representations of $D_{6h}$}

Consider the 2D representation, E$_{2u}$, of hexagonal symmetry, with
strong spin-orbit coupling. The GL functional is constructed from the
amplitudes that parametrize $\vec{f}(\vec{k}_f)$ in terms of the basis
functions of Table I,
\be
\vec{f}(\vec{k}_f) =
\hat{z}\left(\eta_1 \,{\cal Y}_{1}(\vec{k}_f)+
 \eta_2 \,{\cal Y}_{2}(\vec{k}_f)\right)
\,.
\ee
The GL order parameter is then a complex two-component vector,
$\vec{\eta}=(\eta_1,\eta_2)$, transforming according to the E$_{2u}$
representation. The terms in the GL functional must be invariant under
the symmetry group, $G=D_{6h}\times{\cal T}\times U(1)$, of point
rotations, time-reversal and gauge transformations. The form of the GL
functional, $\Omega_{GL}$, is governed by the linearly independent
invariants that can be constructed from fourth-order products,
$\sum\,b_{ijkl}\,\eta_i\eta_j\eta_k^*\eta_l^*$, and second-order
gradient terms, $\sum\,\kappa_{ijkl}(D_i\eta_j)(D_k\eta_l)^*$, where
the gauge-invariant derivatives are denoted by $D_i = \partial_i +
i{2e\over \hbar c} A_i$. The fourth-order product
$\eta_i\eta_j\eta_k^*\eta_l^*$ transforms as $(E_{2}\otimes
E_{2})_{{\rm sym}}\otimes (E_{2}\otimes E_{2})_{{\rm sym}}= (A_1\oplus
E_2)\otimes (A_1\oplus E_2)= 2\,A_1\oplus A_2 \oplus 3\,E_2$, yielding
two linearly independent invariants. Similarly, the in-plane gradient
$D_i\eta_j$ transforms as
$E_1\otimes E_2=B_1\oplus B_2 \oplus E_1$, which generates
three second-order invariants. In addition, the c-axis gradient
$D_z\eta_i$, which transforms as $E_{2}$, yields a fourth
second-order invariant. The resulting GL functional then has the
general form,$^{6,7}$
\ber
\Omega_{GL}\left[{\vec{\;\eta},\vec{A\;}}\right] 
&=&
\int\limits d^{3}R \;\Big\lbrace\,\alpha(T)\vec{\,\eta}\cdot\vec{\eta}^{*}
+\beta_{1}\left({\vec{\,\eta}\cdot\vec{\eta}^{*}}
\right)^{2}+\beta_{2}\left\vert{\vec{\eta}\cdot\vec{\eta}}
\right\vert^{2}
\nonumber\\
&+&
\kappa_{1}
\left({D_{i}\eta_{j}}\right)\left({D_{i}\eta_{j}}\right)^{*}+\kappa_{2}
\left({D_{i}\eta_{i}}\right)\left({D_{j}\eta_{j}}\right)^{*}+\kappa_{3}
\left({D_{i}\eta_{j}}\right)\left({D_{j}\eta_{i}}\right)^{*}
\nonumber\\
&+&
\kappa_{4}
\left({D_{z}\eta_{j}}\right)\left({D_{z}\eta_{j}}\right)^{*}
+\frac{1}{8\pi}\left\{|\vec{\partial}\times\vec{A}|^2
-2\,\vec{H}\cdot\vec{\partial}\times\vec{A}\right\}
\Big\rbrace 
\,.
\eer
The last two terms represent the magnetic field energy for a fixed
external field $\vec{H}$.

The coefficients of each invariant, $\{\alpha (T), \beta_1, \beta_2,
\kappa_1, \kappa_2, \kappa_3, \kappa_4\}$ are material parameters that
must be determined from comparision with experiment or calculated from
a more microscopic theory. A similar analysis follows for any of the 2D
representations, and yields formally equivalent GL functionals, even
though the order parameters belong to different representations. Thus,
at the phenomenological level these GL theories yield identical results
for the thermodynamic and magnetic properties. However, similar GL
theories can differ significantly in their predictions when we
determine the material parameters of the GL functional from a more
fundamental theory, {\ie} the Fermi-liquid theory.

The equilibrium order parameter and current distribution are 
determined by the Euler-Lagrange equations,
\be
{{\delta\Omega_{GL}\left[{\vec{\eta},
\vec{A}}\right]}\over{\delta\eta^{*}_{i}}}=0\,,
\qquad\qquad 
{{\delta\Omega_{GL}\left[{\vec{\eta},\vec{A}}\right]}
\over{\delta A_{i}}}=0\,;
\ee
which yield the GL differential equations for the order parameter,
magnetic field and supercurrent,
\ber
\kappa_{123}D_{x}^{2}\eta_{1}
+\kappa_{1}D_{y}^{2}\eta_{1}
+\kappa_{4}D_{z}^{2}\eta_{1}
+(\kappa_{2}D_{x}D_{y}+\kappa_{3}D_{y}D_{x})\eta_{2}
\nonumber\\
+2\beta_{1}\left({\vec{\eta}\cdot\vec{\eta}^{*}}\right)\eta_{1}
+2\beta_{2}\left({\vec{\eta}\cdot\vec{\eta}}\right)\eta^{*}_{1}
= \alpha\,\eta_{1}
\,,
\eer
\ber
\kappa_{1}D_{x}^{2}\eta_{2}
+\kappa_{123}D_{y}^{2}\eta_{2}
+\kappa_{4}D_{z}^{2}\eta_{2}
+(\kappa_{2}D_{y}D_{x}+\kappa_{3}D_{x}D_{y})\eta_{1}
\nonumber\\
+2\beta_{1}\left({\vec{\eta}\cdot\vec{\eta}^{*}}\right)\eta_{2}
+2\beta_{2}\left({\vec{\eta}\cdot\vec{\eta}}\right)\eta^{*}_{2}
= \alpha\,\eta_{2}
\,,
\eer
and the Maxwell equation,
\ber
(\nabla\times\vec{b})_{i}=
-{{16\pi e}\over{\hbar c}}Im
[\kappa_{1}\,\eta_{j}\left({D_{\perp,i}\eta_{j}}\right)^{*}+
\kappa_{2}\,\eta_{i}\left({D_{\perp,j}\eta_{j}}\right)^{*}+
\kappa_{3}\,\eta_{j}\left({D_{\perp,j}\eta_{i}}\right)^{*}
\nonumber\\
%\hspace{-1.5em}\hskip 3.3in
+\,\kappa_{4}\,\delta_{iz}\eta_{j}\left({D_{z}\eta_{j}}
\right)^{*}]
\,,
\eer
which are the basis for studies of the H-T phase diagram, vortices and
related magnetic properties.$^{22-26}$
I use the notation, $\kappa_{ijk...}=\kappa_i+\kappa_j +\kappa_k +
...$, etc. Below I summarize the basic solutions to the GL theory and
the significance of the material parameters defining the GL
functional.

There are two possible homogeneous equilibrium states depending on the
sign of $\beta_2$. For $-\beta_1<\beta_2<0$ the equilibrium order
parameter, $\vec{\eta}=\eta_0(1,0)$ with $\eta_0 = [|\alpha(T)|
/ 2 \beta_{12}]^{1/2}$, breaks rotational symmetry in the basal plane,
but preserves time-reversal symmetry. The equilibrium state is
rotationally degenerate; however, the degeneracy for an arbitrary
rotations of $\vec{\eta}$ in the basal plane is accidental and is
lifted by higher-order terms in the GL functional. There are three
sixth-order invariants that contribute,
\ber
\delta\Omega_{6}
&=&
\int d^{3}R\,\Big\{
\gamma_{1}\,
\left\vert\vec{\eta}\right\vert^{6}
+
\gamma_{2}\,\left(
\vec{\eta}\cdot\vec{\eta}^{*}\right)\left\vert
\vec{\eta}\cdot\vec{\eta}\right\vert^{2}
\nonumber\\
&+&
\gamma_{3}\,\left[
\left\vert\eta_{1}\right\vert^{6}-
\left\vert\eta_{2}\right\vert^{6}
-3\left(
3\left\vert\eta_{1}\right\vert^{2}\left\vert\eta_{2}\right\vert^{2}
+\eta_{1}^{2}\eta_{2}^{*2}+\eta_{1}^{*2}\eta_{2}^{2}
\right)
\left(\vert\eta_{1}\vert^2 - \vert\eta_{2}\vert^2\right)
\right]\Big\}\,,\quad
\eer
including the leading term in the anisotropy energy, which lifts the
degeneracy and aligns $\vec{\eta}$ along one of six remaining degenerate
directions.

For $\beta_2>0$ the order parameter retains the full rotational
symmetry (provided each rotation is combined with an appropriately
chosen gauge transformation), but spontaneously breaks time-reversal
symmetry. The equilibrium state is doubly-degenerate with an order
parameter of the form $\vec{\eta}_{+}=(\eta_0/\sqrt{2})(1,i)$ [or
$\vec{\eta}_{-}=\vec{\eta}_{+}^{*}$], where $\eta_0 = [|\alpha(T)|
/2 \beta_1]^{1/2}$. The broken time-reversal symmetry of the two
solutions, $\vec{\eta}_{\pm}$, is exhibited by the two possible
orientations of the internal orbital angular momentum,
\be
\vec{M}_{orb}=(\kappa_2 - \kappa_3)\left(\frac{2e}{\hbar c}\right)
\,Im\left({\vec{\eta}\times\vec{\eta}^{\,*}}\right)
\sim\pm\,\hat{c}\;,
\ee
or spontaneous magnetic moment of the Cooper pairs. The presence of
this term in the GL functional is exhibited by rewriting the gradient
terms (for $\vec{H}=0$) in the form,
\ber
\Omega_{grad} 
&=&
\int d^3R\,
\Bigg\{
\kappa_1\,[|\vec{D}_{\perp}\eta_1|^2+|\vec{D}_{\perp}\eta_2|^2]+
\kappa_4\, [|D_z\eta_1|^2+|D_z\eta_2|^2]
\nonumber\\
&+&
\kappa_{23}(|D_x\eta_1|^2+|D_y\eta_2|^2)
\nonumber\\
&+&
\frac{1}{2}\kappa_{23}
\left[(D_x\eta_1)(D_y\eta_2)^* + (D_x\eta_2)(D_y\eta_1)^* +c.c.\right]
\nonumber\\
&+&
(\kappa_2-\kappa_3)\left[\left(\frac{2e}{\hbar c}\right)
(i\vec{\eta}\times\vec{\eta}^*)\cdot(\vec{\partial}\times\vec{A})\right]
\Bigg\}
\,.
\eer

The coefficients of the gradient energy determine the magnitude and
anisotropy of the spatial variations of the order parameter and
supercurrents. The symmetry of the superconducting state depends
critically on the values of the material parameters of the GL
functional. One of the important predictions of Fermi-liquid theory for
the GL free energy of any of the 2D reprepresentations of D$_{6h}$ is
that $\beta_{2}/\beta_{1}\simeq\frac{1}{2}$; thus, the zero-field
equilibrium order paramater spontaneously breaks time-reversal
symmetry. However, estimates of the material parameters from
Fermi-liquid theory predict an orbital moment that is small, and
therefore difficult to observe because of Meissner screening.$^{6,27}$

\section{Fermi-Liquid Theory}

At low temperatures $T\ll T^*$) and low excitation energies
($\epsilon\ll T^*$) the thermodynamic and transport properties of most
strongly interacting Fermion systems are determined by low-lying
excitations obeying Fermi statistics. In metals these excitations,
called `conduction electrons' or `quasiparticles', have charge
$\pm |e|$, spin $\frac{1}{2}$, and even though they are described by
the intrinsic quantum numbers of non-interacting electrons
quasiparticles are complicated states of correlated electrons resulting
from electron-electron, electron-ion, electron-impurity interactions
and Fermi statistics. Fermi-liquid theory has been remarkably sucessful
in describing the low-energy properties of liquid $^3$He and correlated
metals with strong electron-electron and electron-phonon interactions,
including many of the heavy fermions at temperatures below the
coherence temperature $T^{*}$.

The central component of Landau's theory of strongly correlated
Fermions (Fermi-liquid theory) is a classical transport equation
(Boltzmann-Landau transport equation) for the distribution function
$g(\vec{k}_f,\vec{R};\epsilon,t)$ describing the ensemble of
quasiparticle states, where $\vec{k}_f$ is the position on the Fermi
surface, $(\vec{R},t)$ is the space-time coordinates of a quasiparticle
moving with momentum $\vec{k}\simeq\vec{k}_f$ and excitation energy
$\epsilon$.

Green's function techniques have been used to derive the
Boltzmann-Landau transport equation.$^{28-31}$ These methods lead to
expressions for the drift, acceleration and collision terms of
self-energies describing quasiparticle-quasiparticle,
quasiparticle-phonon and quasiparticle-impurity interactions. These
self-energies are functionals of the quasiparticle distribution
function and are defined in terms of interaction vertices between
quasiparticles and other quasiparticle, phonons and impurities. These
interaction vertices have a precise meaning, but their calculation from
first principles is outside the reach of current many-body techniques.
Thus, these interaction vertices are phenomenological material
parameters which must be taken from comparison with experiment. The
most important material parameters, the leading contributions to the
quasiparticle self-energy, are the Fermi surface, Fermi velocity and
density of states at the Fermi energy.

\subsection{The Luttinger-Ward Functional}

In order to derive the Ginzburg-Landau free energy functional, and its
extension to low temperatures, it is useful to formulate an
expression for the thermodynamic potential in terms of the many-body
Green's function. Such a  functional was derived by Luttinger and Ward
for normal fermion systems,$^{10}$ and generalized by deDominicis
and Martin$^{32}$ to superfluid systems.

The starting point is the many-body theory for the one-particle Green's
function. For superfluid Fermi liquids the basic
fermion field must be enlarged in order to describe particle-hole
coherence of the pair condensate.$^{33}$ This is accomplished by introducing
the four-component field operator, $\Psi(x)={\rm
column}(\psi_{\uparrow},\psi_{\downarrow},
\bar{\psi}_{\uparrow},\bar{\psi}_{\downarrow})$, and the $4\times 4$
Nambu Green's function,
\be
\hat{G}_{\mu\nu}(x,x')=
-\left<T_{\tau}\,\Psi_{\mu}(x)\bar{\Psi}_{\nu}(x')\right>
\,, \quad \mu=1,...,4
\,,
\ee
where $x=(\vec{x},\tau)$ denotes the space-imaginary-time coordinate and
$<...>$ is the 
grand ensemble average.
The many-body theory for the Nambu Green's function is derived in the
usual way;$^{34}$ $\hat{G}$ satisfies a matrix Dyson equation with the
$4\times 4$ self energy function $\hat{\Sigma}$ defined in terms of the
skeleton expansion for $\hat\Sigma_{skeleton}[{G}]$.  The particle-hole
space representations of $\hat{\Sigma}$ and $\hat{G}$ are
\be
\hat{G}=\left(
\matrix{
G & F \cr
\bar{F} & \bar{G}
}
\right)
\qquad
\,,
\qquad
\hat{\Sigma}=\left(
\matrix{
\Sigma & \Delta \cr
\bar{\Delta} & \bar{\Sigma}
}
\right)
\,,
\ee
where each element is a $2\times 2$ spin matrix; $G$ is the diagonal
Green's function,
\be
G_{\alpha\beta}(x,x')=
-\left<T_{\tau}\psi_{\alpha}(x)\bar{\psi}_{\beta}(x')\right>
\,,
\ee
and $F$ is the anomalous Green's function,
\be
F_{\alpha\beta}(x,x')=
-\left<T_{\tau}\,\psi_{\alpha}(x)\psi_{\beta}(x')\right>
\,,
\ee
describing the pair-condensate of a superfluid Fermi-liquid.  The
functions $\bar{G}$ and $\bar{F}$ are related by the symmetry
relations, $\bar{G}_{\alpha\beta}(x,x')=-{G}_{\beta\alpha}(x',x)$ and
$\bar{F}_{\alpha\beta}(x,x')={F}_{\beta\alpha}(x'(-\tau'),x(-\tau))^{*}$.

The generalization of the free energy functional of Luttinger and Ward
to superfluid Fermi systems is straight-forward.
Consider a homogeneous system and transform to momentum and frequency
variables: $({x},{x}')\rightarrow(\vec{k},\epsilon_n)$, where $\vec{k}$
is the wavevector and $\epsilon_n=(2n+1)\pi T$ are the Matsubara
frequencies. In this case the generalization of the Luttinger-Ward
free-energy functional is,
\ber
\Omega[\hat{G},\hat{\Sigma}]=-\frac{1}{2}\int\,\frac{d^3k}{(2\pi)^3}\,
&T&
\sum_{n}\,{\rm Tr_{4}}\Big\{\hat{\Sigma}(\vec{k},\epsilon_n)
   \hat{G}(\vec{k},\epsilon_n) 
\nonumber\\
&+& 
\ln[-\hat{G}_0(\vec{k},\epsilon_n)^{-1} + \hat{\Sigma}(\vec{k},\epsilon_n)]
\Big\}
\, + \, \Phi[\hat{G}]\,,
\eer
where $\hat{G}_0(\vec{k},\epsilon_n)$ Green's function for
non-interacting Fermions,
\be
\hat{G}_0(\vec{k},\epsilon_n)^{-1}=
\pmatrix{ i\epsilon_n - \xi^{0}_{\vec{k}} & 0 \cr
           0 & -i\epsilon_n - \xi^{0}_{-\vec{k}} \cr
}\,.
\ee
The log-functional is a formal representation
of the power series in $\hat\Sigma$, and $\Phi[\hat{G}]$ is a
functional which generates the perturbation expansion for the skeleton
self-energy diagrams,
\be
\hat{\Sigma}_{skeleton}[\hat{G}] = 
2\,\frac{\delta\Phi}{\delta \hat{G}^{tr}(\vec{k},\epsilon_n)}
\,.
\ee
Formally, $\Omega$ is a functional of both $\hat{G}$ and $\hat\Sigma$; the
physical Green's function and self-energy are defined by the stationarity
conditions,
\be
\frac{\delta\Omega}{\delta \hat{G}^{tr}(\vec{k},\epsilon_n)}=0
\quad\Rightarrow\quad
\hat{\Sigma}(\vec{k},\epsilon_n)=\hat{\Sigma}_{skeleton}[\hat{G}]
\,,
\ee
\be
\frac{\delta\Omega}{\hat{\Sigma}^{tr}(\vec{k},\epsilon_n)}=0
\quad\Rightarrow\quad
\hat{G}^{-1}(\vec{k},\epsilon_n)=
\hat{G}_0^{-1}(\vec{k},\epsilon_n)-\hat{\Sigma}(\vec{k},\epsilon_n)
\,.
\ee
The first equation identifies $\hat\Sigma$ with the skeleton expansion
evaluated at the physical propagator, while the second stationarity
condition is the Dyson equation. The key point is that $\Omega$ is
equal to the thermodynamic potential when evaluated with $\hat\Sigma$
and $\hat{G}$ that satisfy the stationarity conditions;
$\Omega(T,\mu)=\Omega[\hat{G}_{physical},\hat{\Sigma}_{physical}]$.

\subsection{The Rainer-Serene Functional}

Rainer and Serene$^{11}$ developed this formal machinery into a
powerful calculational scheme for strongly correlated Fermion
superfluids, and used their functional to explain the phase diagram and
thermodynamic properties of superfluid $^3$He. Their formulation of the
free energy functional is based on a classification of the
contributions to the free-energy functional in terms of a set of small
expansion parameters, {\eg} $T/T^{*}$, $T_c/T^{*}$, $1/k_f l$,
$1/k_f\xi_0$, where $T_c$ (pairing energy scale), $l$ (mean free
path) and $\xi_0=v_f/2\pi T_c$ (pair correlation length) represent the
low-energy scales, and $T^{*}$ (degeneracy energy) and $k_f^{-1}$
(Fermi wavelength) are the high-energy scales of the nomal-metal. These
ratios are small in nearly all systems of interest; in most heavy
fermion superconductors $T_c/T^{*}\sim 10^{-1}-10^{-2}$, and in most
metals this ratio is much smaller. The application of the Fermi-liquid
model to the high $T_c$ superconductors is more problematic; however,
special versions of Fermi-liquid theory, {\eg} the nearly
anti-ferromagnetic Fermi-liquid model,$^{35,36}$ and the
coupled-2D-Fermi-liquid model with interlayer diffusion,$^{37}$ are
promising steps towards a theory of superconductivity in layered
cuprates. In the following I develop the
free-energy functional of Rainer and Serene for Fermi-liquid
superconductors and derive the GL free energy functional for an
anisotropic superconductors with an unconventional order parameter.

The propagator-renormalized perturbation expansion for the skeleton
self-\break energy functional (or the $\Phi$-functional) is formulated
in terms of the {\it full} Green's function and {\it bare} vertices for
the electron-electron, electron-ion, and electron-impurity
interactions. Although the skeleton expansion is an exact formulation
of the many-body problem, it is ill-suited for describing the
low-energy properties of strongly correlated Fermions. A formulation of
the many-body perturbation theory in terms of low-energy excitations
(quasiparticles) can be carried out by re-organizing the perturbation
expansion for the self energy (or $\Phi$-functional).

\subsection{High- and low-energy scales}

The idea is to separate $\hat{G}(\vec{k},\epsilon_n)$ into low- and
high-energy parts by introducing a formal scale, $\omega_c$,
intermediate between the high-energy scale (\eg $ T^*$, bandwidth,
etc.) and the relevant low-energy scale, \eg $ T$; \ie $ T \ll \omega_c
< T^{*}$.$^{38,39}$ The non-interacting Green's function is similarly
divided into low- and high-energy parts,
\be
\hat{G}_0(\vec{k},\epsilon_n)=\Bigg\{
\matrix{
        \hat{G}_0(\vec{k},\epsilon_n)_{low} \quad , & 
        |\epsilon_n| < \omega_c \,\, {\rm and} \,\,
|\xi_{\vec{k}}^{0}|<\omega_c \cr
        \hat{G}_0(\vec{k},\epsilon_n)_{high}\quad , &
        |\epsilon_n|\,\,{\rm or}\,\,|\xi_{\vec{k}}^{0}|>\omega_c}
\,.
\ee
%\bigskip\centerline{\tenpoint\bf renormalized vertices and $\Sigma$}\bigskip

Within the low-energy region the bare propagator is of order
$\hat{G}_0(\vec{k},\epsilon_n)_{low}\sim{\sl small}^{-1}$, while the
order of magnitude in the high-energy region is
$\hat{G}_0(\vec{k},\epsilon_n)_{high}\sim{\sl small}^{\, 0}$, where
${\sl small}\sim  T/E_f \ll 1$. Bare vertices (when combined with
appropriate factors of the density of states) are of order ${\sl
small}^{\, 0}$. The perturbation expansion is then enlarged to include
diagrams for both low-energy and high-energy propagators, ad
re-organized into an expansion in terms of low-energy propagators and
{\sl block} vertices.$^{38,39,40}$ The block vertices sum an infinite
set of diagrams composed of bare vertices and high-energy propagators.
An example of a contribution to the self-energy with high- and
low-energy propagators and bare vertices is shown in Fig. (1e); the
self-energy diagram (1d) with the block vertex sums all high-energy
processes that couple to the same topological arrangement of low-energy
propagators.

%%%%%%%%%%%%%%%%%%  Figure 1 %%%%%%%%%%%%%%%%%%%%
\begin{figure}[htbp]
	\centering
		\includegraphics[scale=0.35]{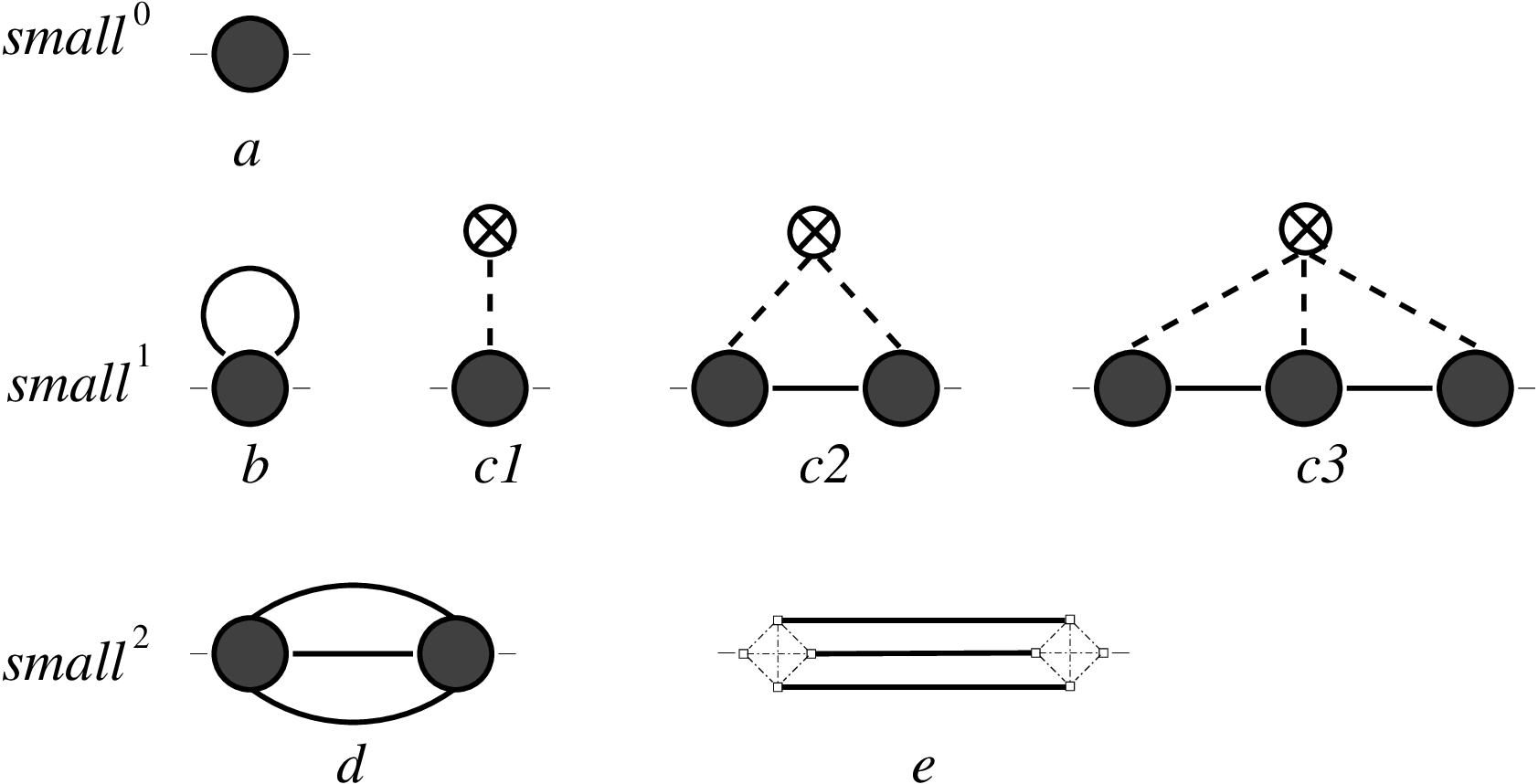}
	\caption{Leading order contributions to the quasiparticle self-energy. Block
		vertices couple to low-energy propagators (solid lines), and represent
		the summation to all orders of the bare interaction (open circle) and
		high-energy intermediate states (thin dotted lines).}
	\label{fig:diagrams}
\end{figure}

%%%%%%%%%%%%%%%%%%%%%%%%%%%%%%%%%%%%%%%%%%%%%

In addition to the summation of high-energy processes into block vertices,
the low-energy propagator is renormalized via the usual Dyson equation,
\be
\hat{G}(\vec{k},\epsilon_n)_{low}=
\hat{G}_0(\vec{k},\epsilon_n)_{low} 
+ 
\hat{G}_0(\vec{k},\epsilon_n)_{low}
\hat{\Sigma}(\vec{k},\epsilon_n)_{low}
\hat{G}(\vec{k},\epsilon_n)_{low}
\,.
\ee
The low-energy part of the self energy can now be defined in terms of
the set of skeleton diagrams defined as the set of $\hat{\Sigma}$ diagrams
in which the self-energy insertions on internal (low-energy) propagator
lines are summed to all orders by the replacement,
$\hat{G}_0(\vec{k},\epsilon_n)_{low} \rightarrow
\hat{G}(\vec{k},\epsilon_n)_{low}$; {\ie}
$\hat{\Sigma}(\vec{k},\epsilon_n)_{low}=
\hat{\Sigma}_{skeleton}[\hat{G}_{low}]$.

The crucial assumption which makes the re-organized many-body
theory\break tractable is that the summation of high-energy processes
into block vertices does not introduce new low-energy physics that is
not accounted for in the low-energy propagator and self-energy. Thus,
all block vertices are estimated to be of order ${\sl small}^{\, 0}$,
and the renormalized low-energy propagator is assumed to be of order
${\sl small}^{-1}$. Furthermore, the block vertices, which depend on
the momenta, $\vec{k}$, and energies, $\epsilon_{n}$, of the initial
and final state (low-energy) excitations, are assumed to vary on the
high-energy scale. Since block vertices couple only to the low-energy
propagators their arguments can be evaluated with
$\vec{k}\simeq\vec{k}_{f}$ and $\epsilon_{n}\simeq 0$. In addition to
the order of magnitude estimates for $\hat{G}_{low}$ and the block
vertices, phase space restrictions, $|\xi_{\vec{k}}|<\omega_c$ and
$|\epsilon_n|<\omega_c$, imply the following factors,
\ber
T\sum_{n}^{|\epsilon_n|<\omega_c}\sim{\sl small}
\quad , \quad
\int_{-\omega_c}^{\omega_c} d\xi_{\vec{k}} \sim{\sl small}
\,,
\nonumber
\eer
for summations and integrations over internal frequencies and momenta.
These power counting rules are subject to constraints imposed by energy
and momentum conservation (for details see Refs. (39)).

\subsection{Leading-order theory}

The leading order contributions to the low-energy electronic
self-energy are shown in Fig.(1a-d). I omit the electron-phonon
contributions; see Refs (39,40). The zeroth-order block vertex is the
contribution to the self-energy from all high-energy processes.  This
term includes bandstructure and correlation effects of ion- and
electron-electron interactions; the quasiparticle residue,
$a(\vec{k}_f)=(1-\partial\Sigma^{(a)}/\partial\epsilon)^{-1}$, is
determined by this term. This zeroth-order self-energy term, including
the quasiparticle residue, is absorbed into the renormalized
quasiparticle dispersion relation, $\xi_{\vec{k}}$, and renormalized
block vertices; \ie $\xi^0_{\vec{k}}\rightarrow
a(\vec{k})(\xi^0_{\vec{k}}-\Sigma^{(a)}(\vec{k}))$, and a factor of
$a(\vec{k})^{1/2}$ is associated with each quasiparticle link to a
block vertex.

The corrections of order ${\sl small}$ are shown as diagrams (b)-(c). Diagram (b), for the terms in particle-hole channel,
corresponds to Landau's Fermi-liquid corrections to the quasiparticle
excitation energy,
\be
\Sigma_{\gamma\alpha}(\vec{k})=\int\,\frac{d^3k'}{(2\pi)^3}\,
  A_{\alpha\beta;\gamma\rho}(\vec{k},\vec{k}')\,
  T\sum_{n'}\,G_{\beta\rho}(\vec{k}',\epsilon_{n'})_{low}
\,,
\ee
while the particle-particle channel contributes the electronic pairing energy,
\be
\Delta_{\alpha\beta}(\vec{k})=\int\,\frac{d^3k'}{(2\pi)^3}\,
  V_{\alpha\beta;\gamma\rho}(\vec{k},\vec{k}')\,
  T\sum_{n'}\,F_{\gamma\rho}(\vec{k}',\epsilon_{n'})_{low}
\,.
\ee

The contributions to the $\Phi$-functional which generate these leading
order diagonal and off-diagonal self-energies are easily constructed from
eqs.(27-28),
\be
\Phi^{(G)}=\frac{1}{4}\,T\sum_{n}\int\,\frac{d^3k}{(2\pi)^3}\,
T\sum_{n'}\int\,\frac{d^3k'}{(2\pi)^3}\,
G_{\gamma\alpha}(\vec{k},\epsilon_n)_{low}\,
A_{\alpha\beta;\gamma\rho}(\vec{k},\vec{k}')\,
G_{\beta\rho}(\vec{k},\epsilon_n)_{low}
\,,
\ee
\be
\Phi^{(F)}=\frac{1}{2}\,T\sum_{n}\int\,\frac{d^3k}{(2\pi)^3}\,
T\sum_{n'}\int\,\frac{d^3k'}{(2\pi)^3}\,
F_{\alpha\beta}(\vec{k},\epsilon_n)_{low}\,
V_{\alpha\beta;\gamma\rho}(\vec{k},\vec{k}')\,
\bar{F}_{\gamma\rho}(\vec{k},\epsilon_n)_{low}
\,.
\ee
The functions, $A_{\alpha\beta;\gamma\rho}(\vec{k},\vec{k}')$, and
$V_{\alpha\beta;\gamma\rho}(\vec{k},\vec{k}')$ represent the block
vertices for the purely electronic interactions in the particle-hole
and particle-particle channels, respectively. These ineractions may be
further separated into spin-scalar and spin-vector functions for the
particle-hole channel,
\be
A_{\alpha\beta;\gamma\rho}(\vec{k},\vec{k}')=
A^{s}(\vec{k},\vec{k}')\,\delta_{\alpha\gamma}\,\delta_{\beta\rho}
+
\vec{\sigma}_{\alpha\gamma}\cdot
{\bf A}^{a}(\vec{k},\vec{k}')\cdot\,
\vec{\sigma}_{\beta\rho}
\,,
\ee
and the spin-singlet and spin-triplet functions for the
particle-particle channels,
\be
V_{\alpha\beta;\gamma\rho}(\vec{k},\vec{k}')=
(i\sigma_y)_{\alpha\beta}\,V^{(e)}(\vec{k},\vec{k}')
\,(i\sigma_y)_{\gamma\rho}
+
\,(i\sigma_y\vec{\sigma})_{\alpha\beta}\cdot
{\bf V}^{(o)}(\vec{k},\vec{k}')\cdot
(i\vec{\sigma}\sigma_y)_{\gamma\rho}
\,.
\ee
Thus, the singlet and triplet components of the pairing self energy become,
\be
\Delta(\vec{k})=\int\, \frac{d^3k'}{(2\pi)^3}\,
V^{(e)}(\vec{k},\vec{k}')\,T\sum_{n'}\,
F(\vec{k}',\epsilon_{n'})_{low}
\,,
\ee
\be
\vec{\Delta}(\vec{k})=\int\, \frac{d^3k'}{(2\pi)^3}\,
{\bf V}^{(o)}(\vec{k},\vec{k}')\cdot\,T\sum_{n'}\,
\vec{F}(\vec{k}',\epsilon_{n'})_{low}
\,,
\ee
where the singlet- and triplet-channel pairing interactions can be expanded
in basis functions defined in terms of the even- and odd-parity irreducible
representations of the crystal point group,
\be
V^{(e)}(\vec{k},\vec{k}')=\sum_{\Gamma}\,V_{\Gamma}\,\sum_{i}^{d_{\Gamma}}\,
{\cal Y}_i^{(\Gamma)}(\vec{k}_f)^{*}
{\cal Y}_i^{(\Gamma)}(\vec{k}_f')
\,,
\ee
\be
{\bf V}^{(o)}(\vec{k},\vec{k}')
=\sum_{\Gamma}\,V_{\Gamma}\,\sum_{i}^{d_{\Gamma}}\,
\vec{{\cal Y}}_i^{(\Gamma)}(\vec{k}_f)^{*}
\otimes
\vec{{\cal Y}}_i^{(\Gamma)}(\vec{k}_f')
\,.
\ee

In addition to these mean-field electronic self-energies, impurity
scattering contributes to leading order in ${\sl small}$. These terms
are first-order in the impurity concentration and are collected to all
orders in the quasiparticle-impurity vertex in the impurity scattering
t-matrix,
\be
\hat{\Sigma}_{imp}(\vec{k},\epsilon_n)=
n_{imp}\,\hat{t}(\vec{k},\vec{k};\epsilon_n)
\,,
\ee
where the low-energy t-matrix is given by,
\be
\hat{t}(\vec{k},\vec{k}';\epsilon_n)=
u(\vec{k},\vec{k}')\,\hat{1} + \int\,\frac{d^3k''}{(2\pi)^3}\,
u(\vec{k},\vec{k}'')\,
\hat{G}(\vec{k}'',\epsilon_n)_{low}\,
\hat{t}(\vec{k}'',\vec{k}';\epsilon_n)\,,
\ee
represents multiple scattering of low-energy excitations by the block
vertex $u(\vec{k},\vec{k}')\,\hat{1}$ for the non-magnetic
quasiparticle-impurity interaction. For weak scattering the t-matrix is
evaluated in second-order in the Born expansion (\ie diagrams (c1) and
(c2)).  The first-order term is absorbed into $\xi_{\vec{k}}$, and the
remaining piece of the impurity self-energy becomes,
\be
\hat{\Sigma}_{imp}(\vec{k},\epsilon_n)= 
\int\,\frac{d^3k'}{(2\pi)^3}\,w(\vec{k},\vec{k}')\,
\hat{G}(\vec{k}',\epsilon_n)_{low}
\,,
\ee
where $w(\vec{k},\vec{k}')=n_{imp}|u(\vec{k},\vec{k}')|^2$ is proportional
to quasiparticle-impurity scattering probability. The corresponding
contribution to the $\Phi$-functional is,
\be
\Phi^{(imp)}=\frac{1}{4}\, T\sum_{n}
\int\,\frac{d^3k}{(2\pi)^3}\,
\int\,\frac{d^3k'}{(2\pi)^3}\,
w(\vec{k},\vec{k}')\,Tr_{4}\,
\{
\hat{G}(\vec{k},\epsilon_n)_{low}
\hat{G}(\vec{k}',\epsilon_n)_{low}
\}
\,.
\ee
Higher order terms in ${\sl small}$ arise from electron-electron
scattering (diagram (d)); this term is responsible for the
inelastic quasiparticle lifetime, $1/\tau_{ee}$, and electronic
strong-coupling corrections to the leading order pairing energies. The
discussion  that follows is confined to the leading order diagrams
which define the Fermi-liquid theory of superconductivity.

The normal-state self energy, $\hat{\Sigma}_N$, and the propagator,
$\hat{G}_N$, are solutions of the stationarity conditions (eqs.(23)-(24)).
To leading order in ${\sl small}$ the normal-state propagator in the
low-energy region is
\be
\frac{1}{a}\hat{G}_N(\vec{k},\epsilon_n)=
\pmatrix{
i\tilde{\epsilon}_n - \xi_{\vec{k}} & 0 \cr
0 & -i\tilde{\epsilon}_n - \xi_{-\vec{k}} \cr
} ^{-1}
\,,
\ee
where $\tilde{\epsilon}_n=\epsilon_n+{\rm sgn}(\epsilon_n)/\tau$, and
$\tau^{-1} = \pi N_f <w(\vec{k}_f,\vec{k}_f'>_{\vec{k}_f'}$ is the
lifetime due to impurity scattering in the normal state. Note that
$<...>_{\vec{k}_f}$ represents the Fermi-surface average and $N_f$ is
the density of states at the Fermi energy.

Since the pairing energy scale is of order ${\it small}$ we can construct a
functional for the
supercondcuting corrections to the normal-metal free energy by subtracting
$\Omega_N=\Omega[\hat{\Sigma}_N,\hat{G}_N]$ from the functional
$\Omega[\hat\Sigma_{low},\hat{G}_{low}]$ to obtain 
\ber
\delta\Omega\left[{\delta\hat{\Sigma},\delta\hat{G}}\right]
=-{{1}\over{2}}T
\sum\limits_{n}\int\limits{{d^{3}k}\over{
({2\pi})^{3}}}
\;Tr_{4}\;\Big\{\delta\hat{\Sigma}({
\vec{k},\epsilon_{n}})\delta\hat{G}
({\vec{k},\epsilon_{n}})\Big\}
\nonumber\\
-{{1}\over{2}}T
\sum\limits_{n}\int\limits{{d^{3}k}\over{({2\pi}
)^{3}}}\;Tr_{4}\;\Big\{\ln'\left[{\hat{1}-\hat{G}_{N}({
\vec{k},\epsilon_{n}})\delta\hat{\Sigma}({
\vec{k},\epsilon_{n}})}\right]\Big\}+
\delta\Phi[{\delta
\hat{G}({\vec{k},\epsilon_{n}})}]\ ,
\eer
where the $\delta\hat{\Sigma}=\hat{\Sigma}-\hat{\Sigma}_{N}$ and
$\delta\hat{G} = \hat{G}-\hat{G}_{N}$ represent the superconducting
corrections to the normal-state self-energy and propagator. Note that
the subtracted functional contains no linear terms in
$\delta\hat{\Sigma}$ or $\delta\hat{G}$ as a result of the stationarity
conditions for the normal state self-energy and propagator. Thus,
$\ln'[1-x]=\ln[1-x] + x$, and the functional $\delta\Phi [\delta\hat{G}]$ is
defined by
\be
\delta\Phi[\delta\hat{G}]=\Phi[\hat{G}_{low}]-\Phi[\hat{G}_N] -
\;Tr_{4}\;\Big\{\frac{\delta\Phi}{\delta\hat{G}^{tr}}\big\vert_{\hat{G}_N}
\,\left(\hat{G}_{low}-\hat{G}_N\right)\Big\}
\,.
\ee

Equation (42) for the superconducting corrections to the low-energy
free energy functional, combined with the renormalized perturbation
expansion in ${\sl small}$, were derived by Rainer and Serene in their
work on the free energy of superfluid $^3$He.$^{11}$ This functional is
used here to derive the Ginzburg-Landau free energy functional and
extensions for several models of unconventional superconductivity.

\subsection{Ginzburg-Landau Expansion}

The free energy functional in eq. (42) depends on both the
self-energies and propagators. In order to derive a free energy
functional which depends only on the order parameter, it is necessary
to eliminate the `excess' information from the full functional.
To leading order in ${\sl small}$ the stationarity condition of
$\delta\Omega$ with respect to the anomalous propagator generates the
mean field self-energy (gap equation) and the impurity renormalization
of the gap function. The gap function provides a convenient definition
of the order parameter.  Reduction of the full functional to a
functional of the order parameter alone is accomplished by inverting
the stationarity condition relating the order parameter,
$\Delta(\vec{k})$, and the anomalous propagator,
$F(\vec{k},\epsilon_n)$.

Since the mean-field self energies are independent of frequency, it is
useful to introduce the frequency-summed propagator,
\be
f_{\alpha\beta}({\vec{k}})=T\sum\limits_{n}
\; F_{\alpha\beta}({\vec{k},\epsilon_{n}})_{low}
\,.
\ee
Equations (33)-(34),
\be
\Delta(\vec{k})=\int\frac{d^3k'}{(2\pi)^3}\;
V^{(e)}(\vec{k},\vec{k}')
f(\vec{k}')
\quad , \quad
\vec{\Delta}(\vec{k})=
\int\,\frac{d^3k'}{(2\pi)^3}\,
{\bf V}^{(o)}(\vec{k},\vec{k}')
\cdot\vec{f}(\vec{k}') 
\,,
\ee
are inverted by expanding the order parameters and propagators in
the basis functions that define the pairing interactions (eq.(3)),
and applying the orthogonality condition,
\be
\int\,d\vec{k}_f\,n(\vec{k}_f)\,
{\cal Y}^{\alpha(\Gamma)}_{i}(\vec{k}_f)
{\cal Y}^{\beta(\Gamma')}_{j}(\vec{k}_f)^*
=\delta_{\Gamma\Gamma'}\,\delta_{\alpha\beta}\,\delta_{ij}
\,,
\ee
which is defined by integration over the Fermi surface with weight
given by $n(\vec{k}_f)$, the angle-resolved density of states at the Fermi
surface; $\int\,d\vec{k}_f\,n(\vec{k}_f)=1$.
The resulting equations are,
\be
\int_{-\omega_c}^{\omega_c} d\xi_{\vec{k}}\,f(\vec{k}) =
\sum_{\Gamma}^{even}\,\frac{1}{N_f V_{\Gamma}}\,
\Delta^{\Gamma}(\vec{k}_f)
\quad , \quad
\int_{-\omega_c}^{\omega_c} d\xi_{\vec{k}}\,\vec{f}(\vec{k}) =
\sum_{\Gamma}^{odd}\,\frac{1}{N_f V_{\Gamma}}\,
\vec{\Delta}^{\Gamma}(\vec{k}_f)
\,,
\ee
where
\be
\Delta^{\Gamma}(\vec{k}_f)=\sum_{i=1}^{d_{\Gamma_{even}}}\,
\eta^{\Gamma}_{i}\,{\cal Y}^{(\Gamma)}_{i}(\vec{k}_f)
\quad , \quad
\vec{\Delta}^{\Gamma}(\vec{k}_f)=\sum_{i=1}^{d_{\Gamma_{odd}}}\,
\eta^{\Gamma}_{i}\,\vec{{\cal Y}}^{(\Gamma)}_{i}(\vec{k}_f)
\,,
\ee
are the order parameters belonging to even- and odd-parity representations
of ${\bf G}$.

The terms in the $\delta\Phi$ functional of order {\sl small} are obtained 
from $\Phi^{(G)}$, $\Phi^{(\bar{G})}$, $\Phi^{(F)}$ and $\Phi^{(imp)}$, and eq.(43) for
$\delta\Phi$. Evaluating
the $\delta\Phi$ functional at the stationary point,
$\delta[\delta\Phi]/\delta[\delta\hat{G}_{low}]=0$, leads to 
\be
\delta\Phi=\frac{1}{4}\,T\sum\limits_{n}
\int\limits{{d^{3}k}\over{\left({2\pi}\right)^{3}}}
\;Tr_{4}\left\lbrace{\delta\hat{\Sigma}(\vec{k},\epsilon_n)
\delta\hat{G}_{low}}(\vec{k},\epsilon_n)
\right\rbrace
\;,
\ee
where $\delta\hat{\Sigma}(\vec{k},\epsilon_n)=\delta\hat{\Sigma}(\vec{k})_{mf}
+\delta\hat{\Sigma}(\vec{k},\epsilon_n)_{imp}$ is determined by eqs.
(27)-(28) and (39). The $\delta\Phi$ contribution to the free energy
functional can then be combined with the first term of eq. (42),
$\delta\Omega_{1}$, to give
$\delta\Omega_{1}+\delta\Phi=\frac{1}{2}\delta\Omega_{1}$. The
contribution to $\delta\Omega_{1}$ coming from the diagonal self energy
and propagator, although formally of order ${\sl small}^{\,2}$,
vanishes at this order; thus,
\be 
\delta\Omega_{1} =
-T\sum\limits_{n}\int\limits{{d^{3}k}\over{
\left({2\pi}\right)^{3}}}\;tr\left\lbrace{\Delta(\vec{k},\epsilon_n)
\bar{F}_{low}(\vec{k},\epsilon_n)} \right\rbrace \,.
\ee

\subsection{Impurities}

Elimination of the low-energy propagator is complicated by the impurity
scattering contribution to $\Delta(\vec{k},\epsilon_n)$. The
probability $w(\vec{k},\vec{k}')$ generally includes scattering in
non-trivial symmetry channels. However, for point impurities (`s-wave'
scatterers), the scattering probability can be replaced by a
angle-independent scattering rate belonging to the identity
representation, $N_f w_0=1/\pi\tau$. Isotropic scattering leads to
considerable simplification in the reduction of
$\delta\Omega[\hat{\Sigma},\hat{G}_{low}]$ to a functional of the
mean-field order parameter. The order parameter is
not renormalized by impurity scattering {\it provided} the scattering
is in a channel other than the pairing channel. Thus, for isotropic
impurity scattering any unconventional order parameter will be
unrenormalized by impurity scattering,
\be
\Delta_{imp}(\vec{k},\epsilon_n) = \left<w(\vec{k}_f,\vec{k}_f')\,
f(\vec{k}_f',\epsilon_n)\right>_{\vec{k}_f'} 
\sim\frac{1}{\tau}\,\left<\Delta(\vec{k}_f')
\,I(\vec{k}_f',\epsilon_n)\right>_{\vec{k}_f'}
\,.
\ee
The stationary condition has been used to write $f(\vec{k},\epsilon_n)=
I(\vec{k},\epsilon_n)\,\Delta(\vec{k}_f)$, where
$I(\vec{k},\epsilon_n)$ is invariant under the symmetry group of the
normal state. Thus, $\Delta_{imp}$ vanishes if $\Delta(\vec{k})$
belongs to any non-identity irreducible representation.  As a result
the impurity terms drop out of eq.(50); eliminating the frequency-summed
propagator then gives,
\be
\frac{1}{2}\delta\Omega_{1}=N_f\,\int\,d\vec{k}_f\, n(\vec{k}_f)\,
\Big\{
\sum_{\Gamma}^{all}\frac{1}{N_f V_{\Gamma}}\,
\frac{1}{2}\,tr\,\left(\Delta^{(\Gamma)}(\vec{k}_f)
 \Delta^{(\Gamma)}(\vec{k}_f)^{\dagger}\right)
\Big\}
\,,
\ee
where $\Delta^{(\Gamma)}(\vec{k}_f)$ is the order parameter belonging
to the representation $\Gamma$.

The remaining terms in $\delta\Omega$ come from the log-functional, and
are evaluated in the GL limit by expanding in
$[\hat{G}_N\hat{\Delta}]$ (again dropping terms higher order than ${\sl
small}^{\,2}$),
\be
\delta\Omega_{ln}=T\sum\limits_{n}\int\limits{{d^{3}k}\over{
\left({2\pi}\right)^{3}}}\sum\limits_{m=1}^{\infty}{{
\left({-1}\right)^{m}}\over{2m}}\left\vert{G_{N}\left({
\vec{k},\epsilon_{n}}\right)}\right\vert^{2m}tr\left\lbrace{
(\Delta\Delta^{\dagger})^{m}}\right\rbrace\;.
\ee
The resulting GL functional expanded through sixth-order in 
$\Delta$ becomes,
\ber
\Omega_{GL}\left[{\Delta}\right]
&=&
\sum\limits_{
\Gamma}\alpha_{\Gamma}\left({T}\right)\left<{{1}\over{2}}tr
\left({\Delta^{(\Gamma)}\Delta^{(\Gamma)\dagger}}\right)\right>+
\beta_{_0}\left<{{1}\over{2}}tr\left({\Delta\Delta^{\dagger}}
\right)^{2}\right>
\nonumber\\
&+&
\gamma_{_0}\left<{{1}\over{2}}tr\left({
\Delta\Delta^{\dagger}}\right)^{3}\right>+\, ... \;,
\eer
where the coefficients are given by,
\be
\alpha_{\Gamma}(T)=N_f\,\Big\{\frac{1}{N_f V_{\Gamma}} -
\pi T\sum_{n}^{|\epsilon_n|<\omega_c}\,
\left(\frac{1}{|\epsilon_n|+1/\tau}\right)\Big\}
\,,
\ee
\be
\beta_{_0}=\frac{1}{4}N_f\,
\pi T\sum_{n}\,
\left(\frac{1}{|\epsilon_n|+1/\tau}\right)^{3}
\quad , \quad
\gamma_{_0}=-\frac{1}{8}N_f\,
\pi T\sum_{n}\,
\left(\frac{1}{|\epsilon_n|+1/\tau}\right)^{5}
\,.
\ee
In the clean limit these parameters reduce to,
\be
\alpha_{\Gamma}(T)=N_f \,\ln(T/T_{c\Gamma})
\;,\;
\beta_{_0}={7
\zeta(3)N_f \over 16\,\pi^{2}T^{2}}
\;,\;
\gamma_{_0}=-{31\,\zeta(5)N_f\over 128\,\pi^4T^{4}}
\,,
\ee
where $T_{c\Gamma}$ is the transition temperature of the $\Gamma$th
irreducible representation; the highest $T_{c\Gamma}$ is the physical
transition temperature $T_c$. In the vicinity of $T_c$, barring a near
degeneracy of two irreducible representations, $\alpha_{\Gamma}$ will
be positive except for the irreducible representation corresponding to
$T_c$; thus, the order parameter will belong to a single representation
near $T_c$.

Non-magnetic impurities are pair-breaking in unconventional
superconductors. The reduction in $T_c$ is contained in eq. (55) for
$\alpha(T)$, which can be written,
\be
\alpha(T)=N_f\,\{\ln(T/T_{c0})+
\psi(\frac{1}{2\pi\tau T}+\frac{1}{2}) -\psi(\frac{1}{2})\}
\,,
\ee
where the pairing interaction $V_{\Gamma}$ and the cutoff, $\omega_c$, have
been eliminated in favor of $T_{c0}$, the clean-limit value for the
transition temperature, and $\psi(x)$ is the di-gamma function. The
equation for $T_c$ is given by the Abrikoso-Gorkov formula,
$\ln(T_{c0}/T_{c})=\psi(\frac{1}{2\pi\tau T_c}+\frac{1}{2})
-\psi(\frac{1}{2})$, but with $\tau$ due to non-magnetic scattering. Near
$T_c$, $\alpha(T)\simeq\alpha'(T-T_c)$; impurity scattering reduces the
coefficient $\alpha'=_f/T_c\{1-\frac{1}{2\pi\tau T_c}\,
\psi'(\frac{1}{2\pi\tau T_c}+\frac{1}{2})\}$. Both $T_c(\tau)$ and
$\alpha'(\tau)$ vanish at the same critical value of $\tau$,
$\frac{1}{2\pi\tau_{cr} T_{c0}}\sim 1$.

The impurity corrections to the higher order GL coefficients can also be
expressed as functions of $\frac{1}{2\pi\tau T_c}$. In particular, the
fourth-order coefficient is given by,
\be
\beta_{_0}=\frac{N_f}{32\pi^2 T_c^2}\,
\left[-\psi''(\frac{1}{2\pi\tau T_c} + \frac{1}{2})\right]
\,,
\ee
which reduces to eq.(57) in the clean limit. In the `dirty' limit,
$\frac{1}{2\pi\tau T_c}\rightarrow\infty$, which onsets rapidly for
$\frac{1}{2\pi\tau T_{c0}}\sim 1$ since $T_c(\tau)$ is strongly
suppressed, $\beta_{_0}^{dirty}\simeq\frac{1}{8}\tau^2$. Thus,
impurities suppress the superconducting transition, but they do not
change the order of the transition.

\subsection{Two-dimensional representations}

In order to proceed further it is necessary to specify the relevant
pairing channel, and the general properties of the pairing interaction,
particularly if an odd-parity channel is involved. The Fermi surface
averages are carried out by expanding $\Delta(\vec{k}_f)$ in the basis
functions of the appropriate irreducible representation. For odd-parity
representations the basis functions, and therefore the spin averages,
depend on the strength of the spin-orbit interactions.

In the absence of spin-orbit coupling, the spin-triplet terms are due
entirely to exchange interactions; thus, the odd-parity vertex is
isotropic under separate rotations of the spin and orbital coordinates of the quasiparticles
linked by the interaction vertex, {\ie}
${\bf V}^{(o)}(\vec{k},\vec{k}')=\,{\bf 1}\,\times\,
\sum_{\Gamma}^{odd}\,V_{\Gamma}\,\sum_{i=1}^{d_{\Gamma}}\,
{\cal Y}_i^{(\Gamma)}(\vec{k}_f)
{\cal Y}_i^{(\Gamma)}(\vec{k}_f')^{*}$. 
The direction of spin of the pairs represents a spontaneously broken
continuous symmetry. Thus, small magnetic fields can orient the spin
components of the order parameter.

By contrast, odd-parity superconductors with strong spin-orbit coupling
have the spin-quantization axis determined (at least in part) by
spin-orbit interactions which are large compared to the pairing energy
scale, {\ie} of order ${\sl small}^{\,0}$. The general form of the
basis functions, $\{ \vec{{\cal Y}}^{(\Gamma)}_{i}(\vec{k}_f) \}$, are
quite complicated. However, if spin-orbit coupling selects a preferred
spin quantization axis along a high-symmetry direction in the crystal,
an `easy axis', then the odd-parity interaction simplifies
considerably.

Consider the case in which the
spin quantization axis is locked to the six-fold rotation axis of a
hexagonal crystal. The odd-parity interaction reduces to,
\be
{\bf V}^{(o)}(\vec{k},\vec{k}')=\,\hat{z}\otimes\hat{z}\,
\sum_{\Gamma}^{odd}\,V_{\Gamma}\,\sum_{i=1}^{d_{\Gamma}}\,
{\cal Y}_i^{(\Gamma)}(\vec{k}_f)
{\cal Y}_i^{(\Gamma)}(\vec{k}_f')^{*}
\,.
\ee
The pairing interaction produces only pairing correlations with
$\vec{\Delta}\sim\hat{z}$, {\ie} odd-parity, $S=1$ pairs with
$\hat{z}\cdot\vec{S}=0$. Thus, strong spin-orbit coupling can have
dramatic effect on the paramagnetic properties of odd-parity
superconductors.

The GL functionals for the 1D representations are formally the same.
The material parameters defining these functionals for different
representations differ by minor factors of order one because of the
Fermi surface averages of the basis functions for higher order
invariants differ slightly. The most significant difference is the
insensitivity of the identity representation to non-magnetic impurity
scattering. For this case the impurity renormalization of $\Delta$ does
not vanish, but cancels the impurity renormalization of the self-energy
in the low-energy propagator $G_N(\vec{k},\epsilon_n)$. All other 1D representations suffer from the pair-breaking
effect of non-magnetic impurities.

The more interesting cases are the 2D representations with strong
spin-orbit coupling.  An order parameter belonging to a 2D
represensentation of $D_{6h}$ has been proposed for the superconducting
phases of UPt$_3$, and an odd-parity representation with
$\vec{\Delta}\sim\hat{z}$ accounts for the
anisotropy of $H_{c2}(T)$ at low temperatures (see Ref. (41) and below).
Consider the odd-parity 2D representation of $D_{6h}$ with the spin
quantization axis locked to $\hat{z}$ by strong spin-orbit coupling. For
either $E_{1u}$ or $E_{2u}$,
\be
\vec{\Delta}(\vec{k}_f)=\hat{z}\,\left(
\eta_{1}\,{\cal Y}_{1}(\vec{k}_f) +
\eta_{2}\,{\cal Y}_{2}(\vec{k}_f)\right)
\,,
\ee
where the orbital functions are listed in Table I. The Fermi-surface
integrals are simplest in the RCP and LCP basis, ${\cal
Y}_{\pm}(\vec{k}_f)\sim\,k_z^{n-1}\,(k_x\pm i k_y)^{n}$ for the
odd-parity $E_{n}$ representation ($n=1,2$).
The Fermi-surface averages for the fourth-order terms lead to the
following:  (i) $<|{\cal Y}_{+}|^4>=<|{\cal Y}_{-}|^4>=
<|{\cal Y}_{+}|^2 |{\cal Y}_{-}|^2>$, (ii) $<{\cal Y}_{+}{\cal
Y}_{-}^{*}(|{\cal Y}_{+}|^2 +|{\cal Y}_{-}|^2)>=0$, and (iii) $<{\cal
Y}_{+}^2 {\cal Y}_{-}^{*2}>=0$. These relations and analogous relations for
the sixth-order terms give the following results or the material parameters
of the GL functional in eq.(8),
\ber
\beta_{1}
&=&
2\beta_{2}=\beta_{_0}<|{\cal Y}_{+}(\vec{k}_f)|^4> 
\nonumber\\
\gamma_{1}
&=&
\frac{2}{3}\gamma_{2}=
 \gamma_{_0}<|{\cal Y}_{+}(\vec{k}_f)|^6> 
\,\,, \quad
\gamma_{3}=0
\,.
\eer

There are a couple of important points to be made here. The sign of
$\beta_2$ determines the relative stability of competing ground states;
$\beta_2>0$ stabalizes a state with broken time-reversal symmetry of
the form $\vec{\eta}\sim(1,\pm i)$. The leading order theory predicts
$\beta_2/\beta_1=\frac{1}{2}$, and therefore a ground state with
broken ${\cal T}$-symmetry for all of the four 2D representations with
strong spin-orbit coupling. The ratio of $\frac{1}{2}$
is independent of the detailed geometry of the Fermi surface, and {\it
insensitive} to impurity scattering. This latter result follows mainly
from the assumption that the scattering probability is dominated by the
channel corresponding to the identity representation, in which case the
impurity renormalization of $\Delta(\vec{k}_f)$ vanishes. The impurity
effects completely factor out of the Fermi surface average when the
scattering rate is isotropic, leaving the ratio for $\beta_2/\beta_1$
to be determined solely by symmetry. The importance of this result is
two-fold: (i) the ground state of any of the 2D models exhibits broken
${\cal T}$ symmetry, and is doubly degenerate with
$\vec{\eta}\sim(1,\pm i)$, and (ii) such a ground state (and therefore
$\beta_2>0$) is a pre-requisite for explaining the double transition in
zero field for UPt$_3$ in terms of a 2D order parameter of $D_{6h}$
coupled to a weak symmetry breaking field. These models are discussed
in detail in Ref. (9). Note also that the hexagonal anisotropy energy,
proportional to $\gamma_3$, vanishes in the leading order
approximation.

The heat capacity jump at $T_c$ for the ground state
$\vec{\eta}_{\pm}\sim\, \left(1,\pm i\right)$ 
is $\Delta C= \alpha ^{2}/ \left( {2 \beta _{1}T_{c}} \right)$, 
which in leading-order theory becomes,
\be
\frac{\Delta C}{\gamma T_c}=\frac{24}{<|{\cal Y}_{+}(\vec{k}_f)|^{4}>}
\,\left\{
\frac{(1-\frac{1}{2\pi\tau T_c}\,
   \psi'(\frac{1}{2\pi\tau T_c}+\frac{1}{2}))^2}
{-\psi''(\frac{1}{2\pi\tau T_c}+\frac{1}{2})}
\right\}
\,,
\ee
which depends on the specific geometry of the Fermi surface 
the details of the basis functions, and the pair-breaking effect of
impurities. In the clean limit, $\Delta C/\gamma T_c = 12\zeta(3)/
<|{\cal Y}_{+}(\vec{k}_f)|^{4}>$. As a rough estimate, the $E_{2u}$ basis
functions with a spherical Fermi surface give a specific heat jump of
$\Delta C/\gamma T_c \simeq 0.97$. Pair-breaking from impurity scattering
leads to further reduction of the heat capacity jump.

Most heavy fermion superconductors show a specific heat jump
significantly different from the value $\Delta C/ \gamma T_{c}= 1.43$
expected for a conventional BCS superconductor. In $UPt_3$ estimates of
$\Delta C/ \gamma T_{c}$ range from $\approx 0.33$ to $\approx
1.0$; more recent reports indicate that this spread in specific
heat jumps is related to sample quality, and in the highest quality
single crystals of $UPt_3$ (highest $T_c$) the specific-heat jump is
$\approx 1.0$.$^{42}$
 
\subsection{Quasiclassical Theory}

The utility and predictive power of Fermi-liquid theory
depends on the plausible, but essential, assumption that the low-energy
propagator is of order ${\sl small}^{-1}$, compared to the high-energy
part of the propagator and the block vertices which are of order ${\sl
small}^{\,0}$, and varies on a scale small compared to the $\omega_c$.
By contrast, the self-energy does not vary rapidly with
$\xi_{\vec{k}}$. This allows a further simplification; the rapid
variations of the low-energy propagator with $\xi_{\vec{k}}$,
corresponding to the fast spatial oscillations of the propagator on the
atomic scale of $k_f^{-1}$, can be integrated out of the low-energy
functional. The relevant low-energy functions are the quasiclassical
propagator,
\be
\hat{g}(\vec{k}_f,\epsilon_n)=\frac{1}{a}\,\int_{-\omega_c}^{\omega_c}
\,d\xi_{\vec{k}_f}\,\hat{\tau}_3\,\hat{G}(\vec{k},\epsilon_n)_{low}
\,,
\ee
and the quasiclassical self-energy,
$\hat{\sigma}(\vec{k}_f,\epsilon_n)=
a\,\hat{\Sigma}(\vec{k}_f,\epsilon_n) \hat{\tau}_3$, where the
particle-hole matrix $\hat{\tau}_3$ is inserted for convenience.
Products of the form $\int\,\frac{d^3k}{(2\pi)^3}\,
\hat{\Sigma}(\vec{k},\epsilon_n)\break
\hat{G}(\vec{k},\epsilon_n)_{low}$ are replaced by Fermi surface averages,
$\int\,d\vec{k}_f\,n(\vec{k}_f)\,
\hat{\sigma}(\vec{k}_f,\epsilon_n)\,\hat{g}(\vec{k}_f,\epsilon_n)$. The
advantage of $\xi_{\vec{k}}$-integrating is that the spatial variations
on length scales large compared to $k_f^{-1}$ are easily incorporated
into the low-energy theory. Such a formulation is essential for the
description of inhomogeneous superconductivity, \eg current carrying
states, vortex structures, Josephson phenomena near interfaces and weak
links, etc.$^{43}$ Indeed many of the unique properties of unconventional
superconductors are connected with spatial variations of the order
parameter on length scales of order $\xi_0=v_f/2\pi T_c$.$^{44}$

Eilenberger's original formulation of the quasiclassical theory starts
from\break Gorkov's equations and integrates out the short-wavelength,
high-energy structure to obtain the transport equation for the
quasiclassical propagator,$^{30}$
\be
\left[i\epsilon_n\hat{\tau}_3 - \hat{\sigma}(\vec{k}_f,\vec{R};\epsilon_n)\,,\,
\hat{g}(\vec{k}_f,\vec{R};\epsilon_n)\right] +
i\vec{v}_f\cdot\vec{\nabla}\hat{g}(\vec{k}_f,\vec{R};\epsilon_n)=0
\,.
\ee
In addition, the quasiclassical propagator satisfies the normalization
condition,
\be
\hat{g}(\vec{k}_f,\vec{R};\epsilon_n)^{2}=-\pi^2\,\hat{1}
\,.
\ee

The generalization of the low-energy free-energy functional to include
spatial variations on scales $R\gg k_f^{-1}$ is straight-forward;
expressed in terms of the quasiclassical propagator and self-energy the
Rainer-Serene functional becomes,
\ber
\delta\Omega\left[{\hat{g},\hat{\sigma}}\right]=
-\frac{1}{2} N_f\int\limits d^{3}R
\int\,d\vec{k}_f\,n(\vec{k}_f)\,T\sum\limits_{n}Tr_{4}
\Big\lbrace\hat{\sigma}({\vec{k}_{f},\vec{R};
\epsilon_{n}})\hat{g}({\vec{k}_{f},\vec{R};
\epsilon_{n}})
\nonumber\\
+\int_{-\omega_c}^{\omega_c}\,d\xi_{\vec{k}}
\;\ln\left[{-\hat{\tilde{G}}_{N}^{-1}({\vec{k},\vec{R};\epsilon_{n}}
)+\hat{\sigma}({\vec{k}_{f},\vec{R};\epsilon_{n}}
)}\right]\Big\rbrace+\delta\Phi\left[{\hat{g}}\right]
\;,
\eer
where
$\hat{\tilde{G}}_N(\vec{k},\vec{R};\epsilon_n)^{-1}
=(i\epsilon_n\hat{\tau}_3-\xi_{\vec{k}}^{op})$
is a differential operator;
$\xi_{\vec{k}}^{op}*g(\vec{k}_f,\vec{R};\epsilon_n)=
\xi_{\vec{k}}\,g(\vec{k}_f,\vec{R};\epsilon_n)-
\frac{i}{2}\vec{v}_f\cdot\vec{\nabla}g(\vec{k}_f,\vec{R};\epsilon_n)$.
As a result the $\xi_{\vec{k}}$-integration is straight-forward only
for homogeneous equilibrium. Nevertheless, this functional provides the
basis for the free-energy analysis of inhomogenous superconducting
states, and for $T\rightarrow T_c$ can be reduced to the GL functional
for spatially varying configurations of the order parameter.

The stationarity condition $\delta[\delta\Omega]/\delta[\hat{\sigma}]=0$
generates the $\xi_{\vec{k}}$-integrated Dyson equation for the low-energy
propagator,
\be
\hat{g}({\vec{k}_{f},\vec{R};\epsilon_{n}})=
\int_{-\omega_c}^{\omega_c} d\xi_{\vec{k}}\,
({\hat{\tilde{G}}_{N}^{-1} - \hat{\sigma}})^{-1}
\;,
\ee
which can be transformed into the quasiclassical transport equation (65).
The transport equation and normalization condition are complete with the
specification of the relevant self energy terms.
In addition to the leading order self-energy terms (Fig. 1),
magnetic fields couple to low-energy quasiparticles to leading order in
${\sl small}$ through the diamagnetic coupling,
\be
\hat{\sigma}_{A}=
\frac{e}{c}\vec{v}_f\cdot\vec{A}(\vec{R})\hat{\tau}_3
\,,
\ee
and the Zeeman coupling,
\be
\hat{\sigma}_{B}=
-B_{i}(\vec{R})\,\mu_{ij}(\vec{k}_f)\,\hat{S}_{j}
\,,
\ee
where $\vec{B}=\vec{\nabla}\times\vec{A}$,
$\hat{\vec{S}}=\frac{1}{2}(\hat{1}+\hat{\tau}_3)\vec{\sigma} +
\frac{1}{2}(\hat{1}-\hat{\tau}_3)\vec{\sigma}^{tr}$ is the
quasiparticle spin operator, and $\mu_{ij}(\vec{k}_f)$ is the effective
moment; for a uniaxial crystal with strong spin-orbit coupling
$\mu_{ij}\rightarrow {\rm diag}\,(\mu_{\perp},\mu_{\perp},\mu_{||})$.
The vertex for the diamagnetic coupling is $\frac{e}{c}\vec{v}_f$, which
follows from gauge invariance.

\subsection{Linearized Gap Equation}

In the following I use the quasiclassical equations to derive the
linearized gap equation for the upper critical field, including the
paramagnetic corrections, and obtain the gradient coefficients of the
GL functional for odd-parity, 2D representations of $D_{6h}$.
Near the second-order transition line ($\Delta/T_c\rightarrow
0$) the quasiclassical equations may be linearized in $\Delta$; the
diagonal propagator is given by its normal-state value,
\be
\hat{g}_N(\vec{k}_f,\vec{R};\epsilon_n)=
-i\pi\,{\rm sgn}(\epsilon_n)\,\hat{\tau}_3
\,,
\ee
while the linear correction is determined by the differential equation,
\be
i\vec{v}_f\cdot\vec{D}f+2i\tilde{\epsilon}_n f +\mu(\vec{\sigma}\cdot\vec{B}\,f+
f\,\sigma_y\vec{\sigma}\sigma_y\cdot\vec{B})=2i\pi\,{\rm sgn}
(\epsilon_n)\,\Delta(\vec{k}_f,\vec{R};\epsilon_n)
\,,
\ee
where the diamagnetic coupling is combined with the derivative, 
$\vec{D}=\vec{\nabla}+i\frac{2e}{c}\vec{A}$ ,
$\tilde{\epsilon}_n=\epsilon_n+{\rm
sgn}(\epsilon_n)/\tau$ includes isotropic impurity scattering, and I assume
an isotropic
effective moment, $\mu$.

For an odd-parity representation the upper critical field is determined
by the the linearized gap equation,
\ber
\vec{\Delta}(\vec{k}_f,\vec{R})=N_f\int d\vec{k}_f'
\,{\bf V}(\vec{k}_f,\vec{k}_f')\cdot\,
2\pi T\sum_{n}\int_{0}^{\infty}ds\exp
\{-2s|\epsilon_n|-{\rm sgn}(\epsilon_n)\,s\,\vec{v}_f'\cdot\vec{D}\}
\nonumber\\
\left[1+\left(\cos(2s\mu B) -1\right)\hat{\bf h}\otimes\hat{\bf h}\right]
\,\vec{\Delta}(\vec{k}_f',\vec{R})
\,,\quad
\eer
where I have specialized to the clean limit, and $\hat{\bf h}$ is the direction of the field. Note that the paramagnetic
effect drops out of the gap equation for any odd-parity representation
with $\vec{\Delta}\perp\vec{B}$. In the case of strong spin-orbit
coupling with the quantization axis locked to $\hat{z}$, \ie ${\bf
V}=\hat{z}\otimes\hat{z}\,V_{\Gamma}\,{\cal
Y}^{(\Gamma)}(\vec{k}_f){\cal Y}^{(\Gamma)}(\vec{k}_f')^{*}$, the
paramagnetic limiting of the upper critical field is strongly
anisotropic, and vanishes for $\vec{B}\perp\hat{z}$. Choi and I
argued that the strong anisotropy of $H_{c2}$ at low temperatures in
UPt$_3$ is evidence for this type of pairing; \ie odd-parity, with
$\vec{\Delta}\sim \hat{z}$ locked by spin-orbit coupling.$^{41}$

The anisotropy of the paramagnetic effect is also observable in the GL
region. The Zeeman coupling contributes a term in the GL functional of the
form,
\be
\Omega_{Zeeman}=g_z\,\int d^3R\,|\vec{B}\cdot\vec{z}|^2\,
\left(|\eta_1|^2+|\eta_2|^2\right) \,,
\ee
where the g-factor is determined by the effective moment,
\be
g_z={7\zeta(3)}\frac{N_f\mu^2}{4\pi^2T_c^2}
\,.
\ee
The important point is that the Zeeman energy is pair-breaking for
$\vec{B}||\hat{z}$, but {\it not} for $\vec{B}\perp\hat{z}$. Thus, if
$\vec{\Delta}||\hat{z}$ is enforced by strong spin-orbit coupling, then
$H^{||}_{c2}(T)$ is suppressed by the Zeeman effect at
low-temperatures, while the temperature dependence of $H_{c2}^{\perp}$
is determined only by diamagnetism; the paramagnetic terms drop out for
this field orientation since the field merely shifts the population of
Cooper pairs with spin directions $|\Leftarrow>$ and $|\Rightarrow>$,
without any loss of condensation energy. The effects of impurity
scattering on both even- and odd-parity superconductors is discussed in
Ref.(45).

\subsection{Gradient Coefficients for the 2D Representations of $D_{6h}$}

Although the phenomenological GL theories are formally the same for any
of the 2D representations, the predictions for the GL material
parameters may differ substantially depending on the geometry of the
Fermi surface and symmetry of the Cooper pair basis functions. For
instance, for the homogeneous terms in the GL functional, the
fourth-order free energy coefficients have the ratio,
$\frac{\beta_2}{\beta_1}=\frac{1}{2}$ to leading order in ${\sl small}$
for any of the four 2D representations; this result is insensitive to
hexagonal anisotropy of the Fermi surface and basis functions, and to
non-magnetic, s-wave impurity scattering.

Significant differences between the 2D representations appear when one
considers the gradient terms in the GL functional. In order
to calculate the leading order gradient terms in the GL equation consider the linearized gap equation (73). Near $T_c$ the estimates
$|\epsilon_n|\sim T_c$, $|\vec{v}_f\cdot\vec{D}|\sim T_c\sqrt{1-T/T_c}$
apply, so that to leading order in $\sqrt{1-T/T_c}$ the linearized
equation for the odd-parity gap function becomes,
\be
\vec{\Delta}(\vec{k}_f,\vec{R})
=\int\,d^2\vec{k}_f'\,n(\vec{k}_f')\,
V(\vec{k}_f,\vec{k}_f')
\left\{{\cal K}(T) +
\frac{7\zeta(3)}{16\pi^2T_c^2}\,(\vec{v}_f'\cdot\vec{D})^2\right\}\,
\vec{\Delta}(\vec{k}_f',\vec{R})
\,,
\ee
where ${\cal K}(T)=\ln(1.13\,\omega_c/T)$. The same equation holds for
the even-parity channel with appropriate substitutions for the gap
function and pairing interaction. This equation is used to generate the
coefficients of the gradient terms in the GL equations.  For any of the
even-parity, or odd-parity (with $\vec{\Delta}||\hat{z}$), 2D models the
gradient coefficients become,
\ber
\kappa_1 &=& \kappa_{0}\,
\left<{\cal Y}_1(\vec{k}_f)\,v_{fy}\,v_{fy}\,{\cal Y}_1(\vec{k}_f)\right> 
\nonumber\\
\kappa_2 &=& \kappa_3=
\kappa_{0}\,
\left<{\cal Y}_1(\vec{k}_f)\,v_{fx}\,v_{fy}\,{\cal Y}_2(\vec{k}_f)\right>
\nonumber\\
\kappa_4 &=& \kappa_{0}\,
\left<{\cal Y}_1(\vec{k}_f)\,v_{fz}\,v_{fz}\,{\cal Y}_1(\vec{k}_f)\right>
\,,
\eer
where ${\cal Y}_i(\vec{k}_f)$ are the basis functions and
$\kappa_{0}=\frac{7\zeta(3)}{16\pi^2 T_c^2}\,N_f$. There are important
differences
between the E$_{1}$ and E$_{2}$ representations when we evaluate
these averages for the in-plane stiffness coefficients.
For a Fermi surface with 
weak hexagonal anisotropy,
$\kappa_2=\kappa_3\simeq\kappa_1$, for the $E_{1}$ representation,
while for the E$_{2}$ representation
\be
\kappa_2=
\kappa_3\ll\kappa_1\sim N_f\left(\frac{v_f^{\perp}}{\pi T_c}\right)^2 
\,.
\ee
In fact, the three in-plane coefficients are identical for the E$_{1}$
model in the limit where the in-plane hexagonal anisotropy of the Fermi
surface vanishes. In contrast, the coefficients $\kappa_2$ and
$\kappa_3$ for the E$_{2}$ model both vanish when the hexagonal
anisotropy of the Fermi surface is neglected. This latter result
follows directly from the approximation of a cylindrically symmetric
Fermi surface and Fermi velocity, $\vec{v}_f =
v_f^{\perp}(\hat{k}_x\,\vec{x} + \hat{k}_y\,\vec{y}) +
v_f^{||}\hat{k}_z\vec{z}$, and the higher angular momentum components
of the E$_{2}$ basis functions, $\kappa_2(E_{2u})
\propto\left<\hat{k}_z(\hat{k}_x^2-\hat{k}_y^2)v_{fx} v_{fy} (2\hat{k}_x
\hat{k}_y)\hat{k}_z\right>\equiv 0$. The importance of Fermi surface
anisotropy on the GL material parameters, and the implications of such 
effects for the identification of the phases of UPt$_3$ is discussed
elsewhere.$^{9,46}$

\par
\medskip
This work was supported by the National Science Foundation through the
Northwestern University Materials Science Center, Grant No. DMR
8821571.
\par

\par
\bigskip

%\noindent
%\begin{pspicture}(0,0)(6,0.5) 
%\psline[linestyle=solid](0,0)(6,0) 
%\end{pspicture} 

%\medskip
\noindent{\bf References}

\begin{enumerate}
\item[1.]
J. Keller, R. Bulla, Th. H\"ohn, and K. Becker, Phys. Rev.
{\bf B41}, ~1878, 1990.

\item[2.]
P. Allen, Comments Cond. Mat. Phys. {\bf 15}, 327, 1992.

\item[3.]
S. Schmitt-Rink, K. Miyake, and C. Varma, Phys. Rev. Lett. {\bf 57}, 2575, 1986.

\item[4.]
L. Taillefer, Physica, {\bf B163}, 278, 1990.

\item[5.]
B. Sarma, M. Levy, S. Adenwalla, and J. Ketterson, Physical Acoustics, {\bf XX}, 107, 1992.

\item[6.]
L. Gor'kov, Sov. Sci. Rev. A ~{\bf 9}, ~1, 1987.

\item[7.]
M. Sigrist and K. Ueda, Rev. Mod. Phys. ~{\bf 63}, ~239, 1991.

\item[8.]
P. Muzikar, D. Rainer, and J. Sauls, NATO-ASI on ``Vortices in Superfluids'', Kluwer Academic Press, 1994.

\item[9.]
J. Sauls, Adv. Phys., 1994 (to appear).

\item[10.]
J. Luttinger and J. Ward, Phys. Rev. ~{\bf 118}, ~1417, 1960.

\item[11.]
D. Rainer  and J. Serene, Phys. Rev. ~{\bf B13}, ~4745, 1976.

\item[12.]
G. Volovik and L. Gor'kov, Phys. JETP ~{\bf 61}, ~843, 1985.

\item[13.]
D. Hess, T. Tokuyasu, and J. Sauls, J. Phys. Condens. Matter, ~{\bf 1},
  ~8135, 1989.

\item[14.]
K. Machida, and M. Ozaki, J. Phys. Soc. Jpn. ~{\bf 58}, ~2244, 1989.

\item[15.]
R. Joynt, V. Mineev, G. Volovik, and M. Zhitomirskii Phys. Rev. ~{\bf B42}, ~2014, 1990.

\item[16.]
K. Machida, and M. Ozaki, Phys. Rev. Lett., ~{\bf 66}, ~3293, 1991.

\item[17.]
M. Palumbo  and P. Muzikar, Phys. Rev.~{\bf B45}, ~12620, 1992.

\item[18.]
D. Chen and A. Garg, Phys. Rev. Lett. ~{\bf 70}, ~1689, 1993.

\item[19.]
M. Zhitomirskii and I. Luk'yanchuk, Sov. Phys. JETP Lett. {\bf 58}, 131, 1993.

\item[20.]
P. Anderson, Phys. Rev. ~{\bf B30}, ~4000, 1984.

\item[21.]
S. Yip  and A. Garg, Phys. Rev. ~{\bf B48}, ~3304, 1993.

\item[22.]
M. Sigrist, T. M. Rice, and K. Ueda, Phys. Rev. Lett.~{\bf 63}, ~1727, 1989.

\item[23.]
T. Tokuyasu, D. Hess, and J. Sauls, Phys. Rev. ~{\bf B41}, ~8891, 1990.

\item[24.]
M. Palumbo, P. Muzikar, J. and Sauls, Phys. Rev. ~{\bf B42}, ~2681, 1990.

\item[25.]
T. Tokuyasu, and J. Sauls, Physica B ~{\bf 165-166}, ~347, 1990.

\item[26.]
Y. Barash and A. S. Mel'nikov, Sov. Phys. JETP ~{\bf 73}, ~170, 1991.

\item[27.]
C. Choi and P. Muzikar, Phys. Rev. ~{\bf B41}, ~1812, 1990.

\item[28.]
G. Eliashberg, Sov. Phys. JETP {\bf 11}, 696, 1960.

\item[29.]
L. Kadanoff and G. Baym, ``Quantum Statistical Mechanics'', W. A. Benjamin, Inc. 1962.

\item[30.]
G. Eilenberger, Z. Physik {\bf 214}, 195 (1968).

\item[31.]
A. Larkin and Y. Ovchinnikov, Sov. Phys. JETP {\bf 28}, 1200, 1969.

\item[32.]
C. De{D}ominicus  and P. Martin, J. Math Phys.~{\bf 5}, ~14, 1964; ~{\bf 5}, ~31, 1964.

\item[33.]
J. Schrieffer, ``Theory of Superconductivity'', W. A. Benjamin, Inc. 1964.

\item[34.]
A. Abrikosov, L. Gor'kov, and I. Dzyaloshinski, ``Methods of Quantum Field Theory in Statistical Physics'', Prentice-Hall, Inc, 1963.

\item[35.]
N. Bulut, D. Hone, D. Scalapino, and N. Bickers, Phys. Rev. {\bf B41}, 1797, 1990.

\item[36.]
A. Millis, H. Monien and D. Pines, Phys. Rev. {\bf B42}, 167, 1990.

\item[37.]
M. Graf, D. Rainer and J. Sauls, Phys. Rev. {\bf B47}, 12087, 1993.

\item[38.]
J. W. Serene and D. Rainer, Phys. Rep.~{\bf 101}, ~221, 1983.

\item[39.]
D. Rainer, in ``Progress in Low Temperature Physics X'', ed. D. F. Brewer, Elsevier Science Pub., Amsterdam, 1986, p. 371.

\item[40.]
D. Rainer and J. Sauls, in ``Strong Coupling Theory of Superconductivity'', Spring College in Condensed Matter Physics on Superconductivity 1992, \break I.C.T.P. Trieste, World Scientific Pub., 1994 (to be published).

\item[41.]
C. Choi and J. Sauls, Phys. Rev. Lett. {\bf 66}, 484, 1991.

\item[42.]
R. Fisher, S. Kim, B. Woodfield, N. Phillips, L. Taillefer, K. Hasselbach,
J. Floquet, A. Giorgi, and J. Smith, Phys. Rev. Lett. ~{\bf 62},~1411, 1989.

\item[43.]
D. Rainer and J. Sauls, Jpn. J. Appl. Phys. {\bf 26}, 1804, 1987.

\item[44.]
C. Choi and P. Muzikar, Phys. Rev. {\bf B39}, 9664, 1989.

\item[45.]
C. Choi and J. Sauls, Phys. Rev. {\bf B48}, 13684, 1993.

\item[46.]
V. Vinokour, J. Sauls, and M. Norman, unpublished.
\end{enumerate}

\end{document}